\documentclass[twocolumn,twocolappendix,fleqn]{aastex701}

\usepackage{enumitem} 
\usepackage{tabularx} 
\usepackage{array} 
\newcolumntype{L}{>{\raggedright\arraybackslash}X} 

\newcommand{\eq}[1]{eq.~(\ref{eq:#1})}

\newcommand{\Eqnb}[1]{Eq. \ref{eq:#1}}
\newcommand{\Eqb}[1]{(Eq.~\ref{eq:#1})}

\newcommand{\se}[1]{Section \ref{sec:#1}}
\newcommand{\app}[1]{Appendix \ref{app:#1}}
\newcommand{\Fig}[1]{Fig.~\ref{fig:#1}}

\newcommand{\Tab}[1]{Table~\ref{tab:#1}}
\newcommand{\be}{\begin{equation}}
\newcommand{\ee}{\end{equation}}
\newcommand{\bad}{\begin{equation} \begin{aligned}}
\newcommand{\ead}{\end{aligned} \end{equation}}

\newcommand{\Msun}{M_\odot}

\newcommand{\g}{\,{\rm g}}

\newcommand{\Mv}{M_{\rm vir}}
\newcommand{\Ms}{M_{\star}}

\newcommand{\Rv}{R_{\rm vir}}

\newcommand{\rhalf}{r_{\rm 1/2}}

\newcommand{\rmax}{r_{\rm max}}

\newcommand{\Vcdm}{V_{\rm circ,DM}}

\newcommand{\Vv}{V_{\rm vir}}

\newcommand{\VmaxDM}{V_{\rm max,DM}}

\newcommand{\jv}{j_{\rm vir}}

\newcommand{\rmd}{{\rm d}}

\usepackage{hyperref}
\usepackage{float}
\usepackage{graphicx}	
\usepackage{amsmath}	
\usepackage{bm}
\usepackage{array}

\newcolumntype{P}[1]{>{\arraybackslash}p{#1}}

\hypersetup{citecolor=blue}

\begin{document}

\title{Connection between galaxy morphology and dark-matter halo structure II: predicting disk structure from dark-matter halo properties}

\author[orcid=0000-0001-8405-2921,sname='Liang']{Jinning Liang}
\affiliation{Department of Astronomy, School of Physics, Peking University, Beijing 100871, China}
\affiliation{Kavli Institute for Astronomy and Astrophysics, Peking University, Beijing 100871, China}
\email[show]{jnliang25@stu.pku.edu.cn}  

\author[orcid=0000-0001-6115-0633,sname='Jiang']{Fangzhou Jiang}
\affiliation{Kavli Institute for Astronomy and Astrophysics, Peking University, Beijing 100871, China}
\email[show]{fangzhou.jiang@pku.edu.cn}

\author[orcid=0000-0001-5356-2419,sname='Mo']{Houjun Mo}
\affiliation{Department of Astronomy, University of Massachusetts, Amherst, MA01003, USA}
\email{hjmo@umass.edu}  

\author[orcid=0000-0001-5501-6008,sname='Benson']{Andrew Benson}
\affiliation{Carnegie Observatories, 813 Santa Barbara Street, Pasadena, CA 91101, USA}
\email{abenson@carnegiescience.edu}  

\author[orcid=0000-0003-3729-1684,sname='Hopkins']{Philip F. Hopkins}
\affiliation{TAPIR, California Institute of Technology, Pasadena, CA 91125, USA}
\email{phopkins@caltech.edu} 

\author[orcid=0000-0003-4174-0374,sname='Dekel']{Avishai Dekel} 
\affiliation{Center for Astrophysics and Planetary Science, Racah Institute of Physics, The Hebrew University, Jerusalem 91904, Israel}
\email{avishai.dekel@mail.huji.ac.il}  

\author[orcid=0000-0001-6947-5846,sname='Ho']{Luis C. Ho}
\affiliation{Kavli Institute for Astronomy and Astrophysics, Peking University, Beijing 100871, China}
\email{lho.pku@gmail.com}  

\correspondingauthor{Fangzhou Jiang}

\begin{abstract}
We investigate how galactic disk structures connect to the detailed properties of their host dark-matter 
halos using the TNG50 simulation. From the hydrodynamic and matched dark-matter–only runs, 
we measure a comprehensive list of halo properties describing density structure, angular momentum, shape, 
assembly history, and environment. Using the morphological decomposition developed in Paper I, we quantify the 
sizes, scale heights, and mass fractions of the disk components for galaxies at $0 \le z \le 4$.
Random Forest (RF) regression shows that halo properties alone predict disk size and thickness with high accuracy, 
while Symbolic Regression (SR) provides compact empirical relations with slightly lower accuracy. 
Disk height is consistently easier to predict than disk size, and lower-mass halos yield higher accuracy 
than massive halos. Predictions based on halo properties measured in the hydro simulations outperform those based on 
halos matched in the dark-matter–only simulation, reflecting the imprint of baryonic restructuring on the inner halo.
SHAP analysis reveals the most informative halo parameters include concentration, Einasto shape, global and inner spin, and recent mass accretion, though their importance varies across disk properties. 
We show correlations between disk size and the density-profile shape arise primarily from disk-induced modification of the inner halo, rather than a primordial connection.
Finally, we point out that disks become more extended with respect to their host halos at higher redshift in low-mass halos, while massive high-redshift halos show the opposite trend. 
We provide SR-based prescriptions that accurately map halo properties to disk structures, offering practical tools for galaxy–halo modeling.
\end{abstract}




\section{Introduction} 

In the $\Lambda$CDM cosmological framework, galaxies form and evolve within dark-matter (DM) halos \citep{White78}, whose mass, shape, and assembly history imprint on their morphologies. 
Yet, despite decades of study, the quantitative link between halo properties and galaxy morphology remains uncertain and incomplete \citep{Mo98,Somerville18,Jiang19,Liang25}.

Stellar disks are particularly revealing, as their sizes \citep[e.g.][]{Kravtsov13,Huang17}, thicknesses \citep[e.g.][]{Zasov02,Sotnikova06}, mass fractions \citep[e.g.][]{Kim13,Romeo20}, and instability \citep[e.g.][]{Romeo23} trace the angular-momentum (AM) acquisition, retention and dynamical heating within halos. 
Previous studies, including Paper I of this series \citep{Liang25}, have shown that, besides halo mass, which is the primary driver of galaxy size, secondary halo properties including spin \citep[e.g.][]{Fall80,Fall83,Mo98}, concentration \citep[e.g.][]{Jiang19}, and accretion rate \citep[e.g.][]{Dubois16} all affect disk sizes, but a comprehensive, multidimensional understanding of how halo conditions shape detailed disk structures is still lacking.

Our goals in this series of studies are to determine how well disk structure can be predicted from halo properties using cosmological simulations, to identify the most influential parameters, and to derive empirical relations that can be incorporated into semi-analytic galaxy-formation models \citep[e.g.][]{Cole00,Lacey16,Henriques20}. 
To this end, we have developed an algorithm that measures the physical morphologies of simulated galaxies by decomposing them into  kinematic components, and we also measure a compressive list of halo structural properties using particle data, which complements the information available from typical halo-finding algorithms (Paper I). 
In this work, the second paper of this series, we make use of these detailed morphological and structural measurements, and apply machine-learning techniques to uncover complex nonlinear relationships in such high-dimensional data. 
We adopt a framework that combines Random Forest regression \citep{Breiman01}, SHapley Additive Explanations \citep{Lundberg17}, and Symbolic Regression \citep[e.g.][]{Koza92} to quantify and interpret the dependence of disk morphology on halo parameters.

The paper is organized as follows. 
\se{Method} describes the simulation data, the measurements of galaxy and halo properties, and the algorithms employed in this work. 
\se{Results} presents the connections between galaxies and dark-matter halos extracted from our measurements, and identifies the most important halo parameters that regulate disk properties.
\se{Discussion} explores the physical reasons for the relations obtained in \se{Results}, and
\se{Conclusion} summarizes our finding. 
Throughout this study, we define halos as spherical over-densities of 200 times the critical density of the Universe, and adopt the cosmological parameters assumed in the TNG simulations.


\section{Method} \label{sec:Method}
\begin{figure*}	
\includegraphics[width=\textwidth]{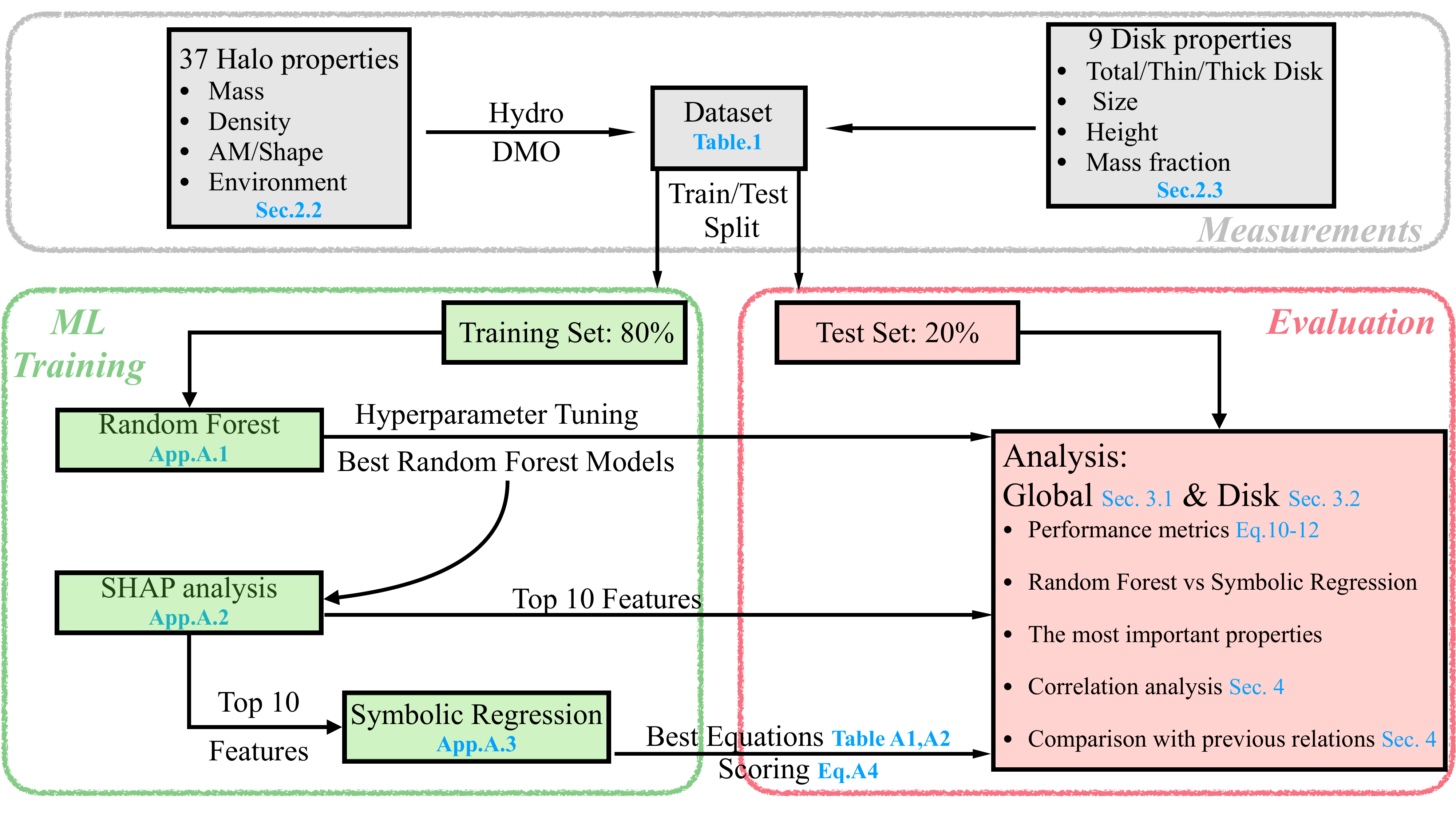}
    \caption{\textbf{Analysis pipeline}, which contains three parts: measurements (gray), machine-learning training (green), and output evaluation (red). The corresponding sections, figures, tables and equations are indicated by blue texts.}
    \label{fig:workflow}
\end{figure*}

In this section, we describe the cosmological simulation sample, define all the halo and galaxy properties considered in this study, and present the machine learning algorithms applied to these data products.
For readers who wish to proceed to the results in \se{Results} without delving into technical details, we provide a summary of our measurements in \Tab{quantities} and a workflow of the analysis pipeline in \Fig{workflow}. 

\subsection{Simulation and sample selection} \label{sec:sim}
We adopt the highest-resolution run of the Illustris-TNG suite \citep[hereafter TNG50,][]{Nelson19,Pillepich19}, which has a gas particle mass of $10^{4.93}\Msun$, a dark matter (DM) particle mass of $10^{5.65}\Msun$, and a gravitational softening length of 0.288 (0.074) comoving kpc for collisionless (gas) particles. 
The simulation includes sub-grid models that capture the key physical processes of gas cooling, star formation, chemical enrichment, and feedback. Halos are identified using the Friends-of-Friends algorithm \citep[FoF,][]{Davis85} and the {\tt SUBFIND} algorithm \citep{Springel01}, and are linked across snapshots via the {\tt SUBLINK} merger tree algorithm \citep{Rodriguez-Gomez15}. The corresponding DM-only (DMO) simulation shares the same initial conditions, and halos are matched to their hydro-simulation counterparts using the public {\tt Subhalo Matching To Dark} catalog\footnote{\href{https://www.tng-project.org/data/docs/specifications/\#sec5d}{https://www.tng-project.org/data/docs/specifications/\#sec5d}}.

Our sample selection follows Paper I and includes central halos at redshift $z=0$, 1, 2, and 4 in the public catalog\footnote{\href{https://www.tng-project.org/data/downloads/TNG50-1}{https://www.tng-project.org/data/downloads/TNG50-1}}. 
We restrict our analysis to well resolved systems, by requiring a half-stellar-mass radius at least twice the collisionless softening length, and at least 1000 stellar particles and 1000 DM particles within the virial radius. 
For the DM-only runs, we only include halos with matched hydro-simulation counterparts. 

\begin{table*}[ht!]
\centering
\renewcommand{\arraystretch}{1}
\begin{tabularx}{0.99\textwidth}{l L} 
\hline
\hline
Halo quantities & Description  \\ 
\hline
$\Mv$ ($\Rv$) & Mass (radius) of a spherical overdensity that is 200 times the critical density \\
$c$ & Halo concentration, measured by fitting Einasto profile \Eqb{Einasto}  \\
$\alpha$ & Halo profile-shape index, measured by fitting Einasto profile \Eqb{Einasto} \\
$\lambda$ ($\lambda_{\rm inner}$)& The spin parameter \Eqb{spin}, measured at $R_{\rm vir}$ ($0.1 R_{\rm vir}$) \\
$q$ ($q_{\rm inner}$) & Axis ratio between intermediate axis and major axis \Eqb{axisratio}, measured from the 3D inertia tensor within $R_{\rm vir}$ ($0.1 R_{\rm vir}$) \\
$p$ ($p_{\rm inner}$) & Axis ratio between minor axis and intermediate axis  \\
$s$ ($s_{\rm inner}$) & Axis ratio between minor axis and major axis \\
$T$ ($T_{\rm inner}$) & Triaxiality, \Eqb{triaxiality}, measured within $R_{\rm vir}$ ($0.1 R_{\rm vir}$) \\
$\dot{M}_{\rm norm}$ & The normalized accretion rate \Eqb{normacc} \\
PC1 & The first principal component of the mass assembly history \\
PC2 & The second principal component of the mass assembly history \\
$z_{1/2}$ & The highest redshift when the main branch of a halo assembled half of its final mass \\
Environment & Cosmic web classification, which categorizes halos as voids, sheets, filaments, or knots using eigenvalues of the deformation tensor \\
$n_{1,3,5}$ ($\delta_{1,3,5}$) & Local number density (overdensity) of all the halos with mass exceeding 0.1\% of the target halo mass \Eqb{localdensities}, measured within 1, 3, or 5 Mpc \\
$z_{\rm Minor/Major/Total}$ & Redshift of the last merger within the latest four dynamical times, for minor, major, or total mergers \\
$\langle z \rangle_{\rm Minor/Major/Total}$ & Mean redshift of mergers within the latest four dynamical times \\ 
$\langle R \rangle_{\rm Minor/Major/Total}$  & Mean mass ratio for the mergers within the latest four dynamical times  \\
$N_{\rm Minor/Major/Total}$ & Number of mergers within the latest four dynamical times \\
\hline
Galaxy quantities  & Description\\ 
\hline
$M_\star$ & Stellar mass (measured within 5$r_{1/2}$, excluding wind particles)  \\
SFR & Star formation rate (sum of the star formation rates of all the gas cells in the halo) \\
$r_{1/2}$  & 3D radius within which the stellar mass equals half of the total stellar mass in the halo \\
$R_{\rm 1/2, Disk/ThinDisk/ThickDisk}$ &  Projected radius enclosing half the mass of the total/thin/thick disk \\
$Z_{\rm 1/2, Disk/ThinDisk/ThickDisk}$ & Vertical distance enclosing half the total/thin/thick disk mass  \\
$f_{\rm Disk/ThinDisk/ThickDisk}$ & Mass fraction of the galaxy in the total/thin/thick disk \\
\hline
\end{tabularx}
\caption{Structural and environmental properties of DM halos and morphological measurements of galaxies used in this study.}\label{tab:quantities}
\end{table*}

\subsection{DM halo measurements}\label{sec:halo}

\begin{itemize}[leftmargin=*]

\item \textbf{Halo mass $\Mv$} is defined as the total mass enclosed within the \textbf{virial radius $\Rv$}, inside which the mean density is $\Delta_{\rm c}=200$ times the critical density $\rho_{\rm c}=3H_0^2/(8\pi G)$, i.e.
$M_{\rm 200}=4\pi \Delta_{\rm c}\rho_{\rm c} R_{\rm 200}^3/3$.
We use the fields of \texttt{Group\_R\_Crit200} and \texttt{Group\_M\_Crit200} from the public halo catalog for $\Rv$ and $\Mv$, respectively.

\item \textbf{Halo circular velocity} is defined as $\Vcdm=\sqrt{G M_{\rm DM}(<r)/r}$, where $M_{\rm DM}(<r)$ is the enclosed DM mass. 
Its maximum value $\VmaxDM$ and the corresponding radius $\rmax$ satisfy $\Vcdm(\rmax)=\VmaxDM$.

\item \textbf{Halo concentration} is defined as $c=\Rv/r_{-2}$, where $r_{-2}$ is the radius where the density slope equals –2.
We fit the \citet{Einasto65} profile
\be\label{eq:Einasto}
\rho(r)=\rho_{-2}\exp\!\left\{-\frac{2}{\alpha}\!\left[\!\left(\frac{r}{r_{-2}}\right)^{\alpha}-1\!\right]\right\},
\ee
for concentration and the \textbf{shape index $\alpha$}.
The normalization $\rho_{-2}$ is linked to $\Mv$ through
\[
\rho_{-2}=\frac{\Mv\alpha}{4\pi h^3 e^{2/\alpha}\gamma[3/\alpha,x(\Rv)]},
\]
with $h=r_{-2}(\alpha/2)^{1/\alpha}$, $x=(2/\alpha)(r/r_{-2})^{\alpha}$, and $\gamma(a,x)$ the non-normalized incomplete gamma function.
Both $c$ and $\alpha$ are obtained by fitting this model to the $\Vcdm/\VmaxDM$ profile.

\item \textbf{Halo spin parameter} is defined as
\be\label{eq:spin}
\lambda=\frac{\jv}{\sqrt{2}\Rv\Vv},
\ee
following \citet{Bullock01}, where $\jv$ is the specific angular momentum within the virial radius and $\Vv$ the circular velocity at $\Rv$.
We also measure an inner spin,
$\lambda_{\rm inner} = j_{\rm vir, inner} / (\sqrt{2}\Rv\Vv)$,
using the specific angular momentum enclosed within $0.1\Rv$, $j_{\rm vir, inner} $.

\item \textbf{The mass accretion rate and mass assembly history}
(MAH) of a dark-matter halo are characterized using two approaches.
The MAH is defined as the mass of the main branch as a function of redshift, $M_{\rm vir}(z)$. 
Following \citet{McBride09}, we fit it with 
\be\label{eq:MAH}
M_{\rm vir}(z) = M_{\rm vir}\,(1+z)^{\beta} e^{-\gamma z}.
\ee
The accretion rate can be defined as
\be\label{eq:normacc}
\dot M_{\rm norm} \equiv \frac{\rmd\ln M_{\rm vir}}{\rmd z} = \frac{\beta}{1+z} - \gamma .
\ee
Alternatively, the MAH can be expressed using the mass variable \citep{vdBosch02,Wechsler02,Zhao03},
\be\label{eq:massvariable}
s(z) = \frac{\sigma(M_{\rm vir})}{\sigma[M_{\rm vir}(z)]},
\ee
where $\sigma(M)$ is the standard deviation of the linear over-density field on mass scale $M$.
Following \citet{Chen20}, we apply Principal Component Analysis (PCA) to $s(z)$ at fixed redshift to reduce dimensionality. 
The first two principal components (PC1 and PC2) accurately capture both the shape and evolution of a typical MAH, with PC1 representing the overall monotonic growth, while PC2 encodeing a higher-order, non-monotonic deviation.
Thus, we use $\dot M_{\rm norm}$ to trace recent accretion and (PC1, PC2) to describe the global MAH shape over cosmic time.

\item \textbf{Halo formation redshift} $z_{\rm 1/2}$ is defined as the highest redshift at which the main branch assembled half of its final mass $M_{\rm vir}$ \citep{Wang11}.

\item \textbf{Halo 3D shape parameters} are derived from the eigenvalues of the inertia tensor \citep{Allgood06},
\be\label{eq:inertia}
\mathcal{S}_{ij} = \frac{1}{M}\sum_k m_k\, r_{k,i}\, r_{k,j},
\ee
where the sum runs over all DM particles within the ellipsoid, and $M=\sum_k m_k$ is the enclosed mass.
The eigenvalues of $\mathcal{S}$ correspond to the squared semi-axes ($a \ge b \ge c$), obtained iteratively from the virial sphere until convergence \citep{Tomassetti16}.
We define the 3D axis ratios
\be\label{eq:axisratio}
q=b/a,\quad p=c/b,\quad s=c/a,
\ee
and the triaxiality parameter
\be\label{eq:triaxiality}
T = (1 - q^2)/(1 - s^2).
\ee
Analogous quantities measured within $0.1R_{\rm vir}$ are denoted as $q_{\rm inner}$, $p_{\rm inner}$, $s_{\rm inner}$, and $T_{\rm inner}$.

\item \textbf{Environment classification} is based on a halo’s location within the cosmic web.
We identify web types using the eigenvalues of the deformation tensor \citep[e.g.,][]{Hahn07,Forero-Romero09}: the number of eigenvalues exceeding a threshold $\lambda_{\rm th}=0.4$ determines the environment, with 0, 1, 2, and 3 corresponding to voids, sheets, filaments, and knots, respectively.
The overdensity field is computed on a $512^3$ grid via a cloud-in-cell interpolation, smoothed with a Gaussian filter of $R_{\rm G}=0.5\,h^{-1}{\rm ckpc}$.
These parameter choices yield a visually consistent web morphology and agree with previous studies \citep[e.g.,][]{Martizzi19}.

\item \textbf{Local densities} quantify a halo’s immediate environment.
We measure both the number density $n_{1,3,5}$ and overdensity $\delta_{1,3,5}$ within spheres of radius $R =$ 1, 3, and 5 Mpc.
Only neighboring halos with total mass $>0.1\%$ of the target’s mass are included:
\be\label{eq:localdensities}
n_R = \sum_{D_i<R} \frac{3}{4\pi R^3}, \qquad
\delta_R = \frac{1}{\rho_{\rm m}}\sum_{D_i<R}\frac{3M_i}{4\pi R^3} - 1,
\ee
where $D_i$ is the distance and $M_i$ the mass of each neighbor, and $\rho_{\rm m}$ is the mean cosmic matter density at that redshift.

\item \textbf{Merger statistics} are extracted from the \texttt{SUBLINK} merger trees.
To ensure relevance to instantaneous properties, only mergers within the latest four dynamical times are considered, where the halo's dynamical time is given by $t_{\rm dyn}=1/[10H(z)]$, with $H(z)$ the Hubble constant at redshift $z$.
We include events where the secondary progenitor contains $\ge 50$ DM particles, and define the merger mass ratio $R$ as the mass ratio of the secondary halo to the
primary, evaluated at the epoch when the secondary reaches its maximum DM mass.
Mergers are classified as major if $R>0.25$ or minor if $0.1<R<0.25$; smaller ratios are ignored.
For each halo, we record the most recent merger redshift ($z_{\rm Minor/Major/Total}$), mean mass ratio ($\langle R\rangle_{\rm Minor/Major/Total}$), number of mergers ($N_{\rm Minor/Major/Total}$), and mean merger redshift ($\langle z\rangle_{\rm Minor/Major/Total}$) for minor, major, and all mergers.

\end{itemize}

\subsection{Galaxy Quantities}\label{sec:galaxy}

\begin{itemize}[leftmargin=*]
\item \textbf{Half-stellar-mass radius} $\rhalf$ is the 3D radius within which the enclosed stellar mass is equal to half of the total stellar mass within the halo.

\item \textbf{Stellar mass} $\Ms$ is defined as the sum of stellar particle mass within 5$\rhalf$. 
This excludes most of satellite galaxies while retaining most of the smooth stellar halo of the central galaxies. 

\item \textbf{Star formation rate} (SFR) is the sum of the star formation rates of all the gas cells in a halo, as in the \texttt{SubhaloSFR} field of the public {\tt SUBFIND} catalog.

\item \textbf{Disk mass fraction} is calculated using the kinematic morphological decomposition algorithm introduced in Paper I\footnote{\href{https://github.com/JinningLianggithub/MorphDecom}{https://github.com/JinningLianggithub/MorphDecom}}. 
This method splits the stellar particles in a galaxy into bulge, stellar halo, thin disk, and thick disk, using automatically detected thresholds in the energy-angular-momentum space.
Disk mass fraction is then defined as $f_{\rm X}=M_{\star,X}/\Ms$ where $M_{\star,X}$ is the mass of component $X$, with $X$ being the total disk, thin disk, or thick disk.

\item \textbf{Scale radius $R_{\rm disk}$ and scale height $Z_{\rm disk}$ of disks} are defined as the projected radius or height enclosing half of the stellar mass of the corresponding component (total disk, thin disk, or thick disk), computed in the coordinate frame whose vertical axis is aligned with the galaxy's angular-momentum vector. 
\end{itemize}

\subsection{Machine Learning Algorithms}\label{sec:ML}

While (semi-)analytical and empirical models have used connections between disk size and halo spin \citep[e.g.][]{Cole00,Lacey16,Henriques20} or concentration \citep[e.g.][]{Nadler20,Behroozi22,Somerville25}, it remains an open question which one is more influential \citep{Jiang19} and whether incorporating a broader range of halo properties can enhance prediction accuracy \citep{Liang25}.
Moreover, it is not yet known whether disk vertical height and mass fraction, can be robustly inferred from halo properties. Addressing these issues necessitates a systematic investigation of correlations across a large set of halo conditions, a task that can be effectively done by supervised machine-learning (ML) algorithms.

In this study, we use ML algorithms to uncover relations between disk parameters and halo conditions. 
The targets and features are galaxy quantities and halo quantities, respectively, which are summarized in \Tab{quantities}. 
We consider two ML algorithms. 
We first train the random forest (RF) regressor \citep{Breiman01} for all halo conditions to predict our targets. 
Then we use SHapley Additive Explanations (SHAP) values \citep{Lundberg17} to identify important halo properties that affect each prediction. 
Finally we use symbolic regression \citep[SR, e.g.][]{Koza92} implemented in the package \texttt{PySR} \citep{Cranmer23}, with the most important 10 halo properties as identified by the RF regressor as input to derive empirical functions relating the targets and features. 
The principle of ML algorithms and hyper-parameters that we tune are described in \app{MLalgorithm}. The workflow in this work is presented in \Fig{workflow}.

For training SR models, we adopt a loss function $\mathcal{L}$ that combines a normalized mean squared error term $\mathcal{L}_{\rm NMSE}$ and a normalized quantile loss term $\mathcal{L}_{\rm NQL}$,
\be
\mathcal{L}=\mathcal{L}_{\rm NMSE}+\mathcal{L}_{\rm NQL}.
\ee
Here, the normalized mean squared error term takes form
\be\label{eq:NMSE}
\mathcal{L}_{\rm NMSE} = (y-\hat{y})^2/y^2,
\ee
where $y$ is the data and $\hat{y}$ is the predicted values. 
The normalized quantile loss is given by
\begin{align}\label{eq:NQL}
\mathcal{L}_{\rm NQL} 
&= \frac{1}{5} \sum_{i=1}^{\tau_i} 
\Bigg[ \tau_i \max\Big(\frac{y-\hat{y}}{y}, 0\Big) \notag \\
&\quad + (1-\tau_i) \max\Big(-\frac{y-\hat{y}}{y}, 0\Big) \Bigg],
\end{align}
which captures the asymmetric penalties for underestimation and overestimation across multiple quantiles $\tau_i$ (0.1, 0.25, 0.5, 0.75, and 0.9). 
This loss function is able to capture information about the entire conditional distribution of the target variable, rather than only the mean. 
This is particularly useful when the data contain outliers or are asymmetrically distributed. 
At early training stage, when the residuals are large, the NMSE term typically dominates and encourages fit of the mean values.  
As training progresses and residuals become small, the NMSE term decreases, while the NQL term become comparable or more influential, guiding the model towards improved distribution match, avoiding many outliers in the predictions. 
This loss-function design is adopted in previous astrophysical studies \citep[e.g.][]{Salim25}, and theoretical considerations can be found in \cite{Terven25}.

Besides the model expressions, \texttt{PySR} also returns the loss $\mathcal{L}$ and complexity $\mathcal{C}$. 
More complex expressions typically achieve lower loss but are more difficult to implement into semi-analytical models. 
To balance complexity and loss, we devise a score function to automatically select equations. 
We first construct normalized loss $\tilde{\mathcal{L}}$ and normalized complexity $\tilde{\mathcal{C}}$ as
\be
\tilde{\mathcal{L}}=\frac{\mathcal{L}-\min(\mathcal{L})}{\max(\mathcal{L})-\min(\mathcal{L})},\ \tilde{\mathcal{C}}=\frac{\mathcal{C}-\min(\mathcal{C})}{\max(\mathcal{C})-\min(\mathcal{C})},
\ee
respectively.
Then the score is given by
\be\label{eq:score}
\text{Score}=\Bigg[w\log(1+\tilde{\mathcal{L}})+(1-w)\log(1+\tilde{\mathcal{C}})\Bigg]^{-1},
\ee
where $w$ is the weight. 
This formula first increases with complexity, reaching a maximum, and then decreases with complexity. 
We choose the equation at the maximum point. 
We find that the loss decreases very slowly once the model complexity exceeds a certain threshold value. 
To identify the simplest equation whose complexity is just sufficient to reach this value, we find, by trial and error, a complexity weight of $w=0.7$ is optimal. 
A higher $w$ results in greater complexity but lower loss at the maximum point.

To avoid overly complicated SR relations and overfitting, we restrict the allowed operators to multiplication ($\times$), plus ($+$), power (\ $ \widehat{} $\ ), exponential ($\exp$), and base-10 logarithm ($\log$). The maximum expression complexity is set to 100. Further details on the SR parameters are provided in \app{SR}.

For both RF and SR, we use 80\% of the total samples as the training set and the remaining 20\% as the test set. 
All features and target variables are scaled to dimensionless quantities. 
For the targets such as disk scale radii, disk scale heights, and galaxy sizes, they are normalized by the virial radius, and the dimensionless quantities are referred to as \textit{compactness} or \textit{thickness}. 
The stellar mass is normalized by the virial mass, while the SFR is normalized by the average halo growth rate over Hubble time. 
For the features, the virial mass is scaled by $10^{10}\Msun$, $M_{\rm vir,10}=M_{\rm vir}/10^{10}{\rm M}_\odot$, and the other properties are already dimensionless. 
Except for disk mass fractions, a base-10 logarithm is applied to the target variables.

\begin{figure*}
\centering
\includegraphics[width=0.86\textwidth]{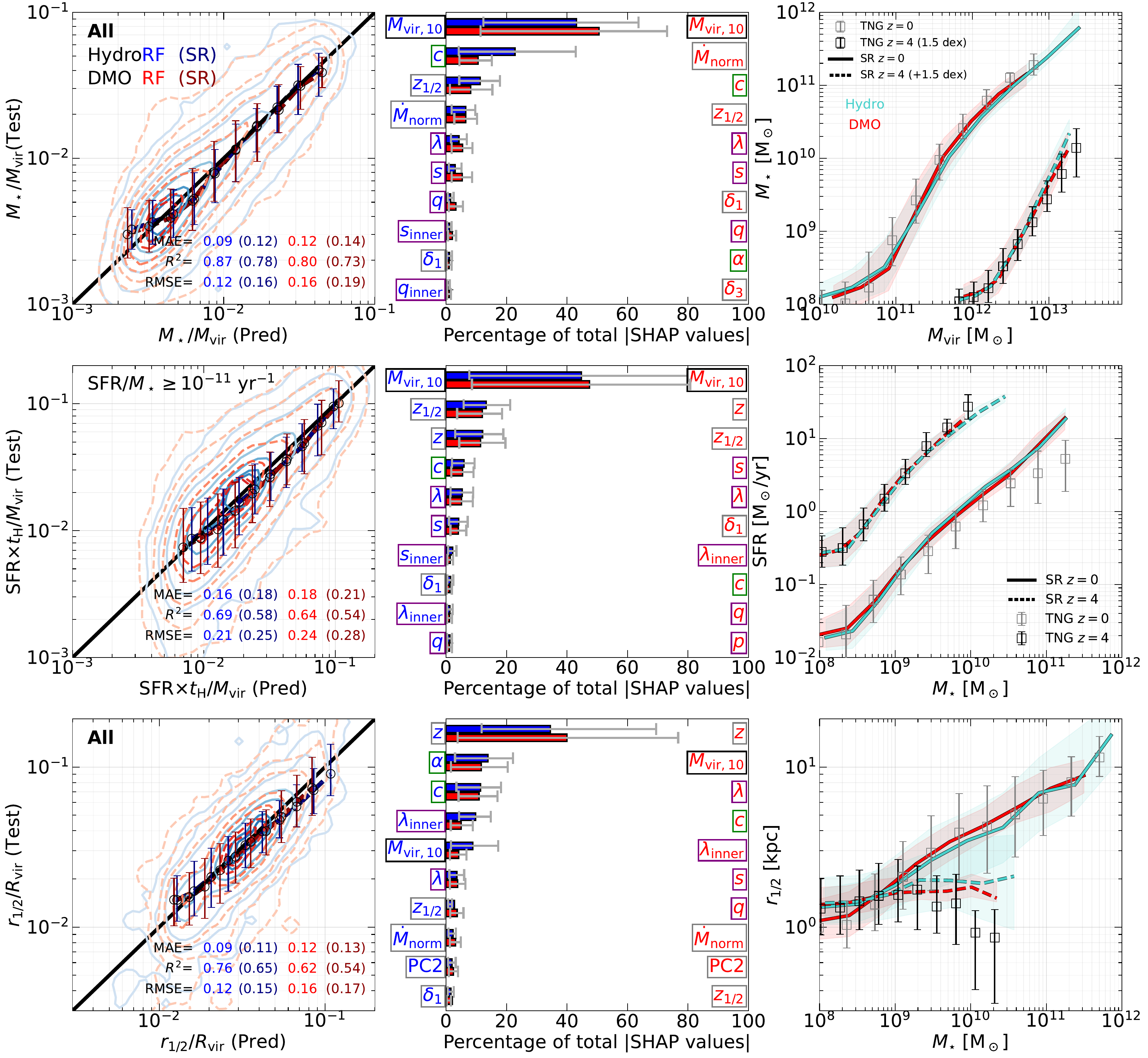}
    \caption{
\textbf{Performance of the regression models for global galaxy properties in the TNG50 simulation, and the scaling relations predicted by the symbolic-regression (SR) models.}
The global galaxy properties shown are stellar mass ({\it top}), SFR ({\it middle}), and half-stellar-mass radius ({\it bottom}).
\quad
{\it Left}: Comparison of RF (contours) and SR (dashed curves with 16–84\% ranges) predictions against test data, using measurements from the hydro (blue) and DMO (red) runs. Performance is evaluated with MAE, $R^2$, and RMSE.
\quad
{\it Middle}: SHAP scores that indicate the relative contributions of the top 10 RF features, with blue (left axis) and red (right axis) labels corresponding to the hydro and DMO results, respectively. 
Error bars show the 16–84\% ranges. Feature categories are indicated by the color of the text boxes: mass (black), density profile (green), angular momentum or trivial shape (purple), and environment (gray).
\quad
{\it Right}: Scaling relations predicted by SR, where cyan and red curves show median relations for hydro halos and DMO halos, respectively, at $z=0$ (solid) and $z=4$ (dashed). 
Gray and black squares show the corresponding medians measured directly from the hydro simulation at $z=0$ and $z=4$. 
Shaded regions and error bars denote 16–84\% ranges.
\quad
The sample in the SFR row includes only star-forming galaxies. For clarity, the $z=4$ relations in the $\Ms$–$\Mv$ panel are shifted horizontally by 1.5 dex.
} 
\label{fig:validation}
\end{figure*}

We emphasize and caution that with SR, we do not aim at achieving a physically interpretable expressions, but instead, we aim at accurate and practical formula for predicting galaxy properties using purely halo measurements.
We experiment with the complexity of the formula to achieve minimal set of quantities and low complexity while keeping a decent accuracy comparable to the full RF models.

To evaluate the performance of the functions derived by RF and SR, we use three different metrics defined as follows.
First, the mean absolute error (MAE), 
\be
\text{MAE}=\frac{1}{N}\sum_{i}^{N}|y_{{\rm true},i}-y_{{\rm pred},i}|.
\ee
Second, the coefficient of determination ($R^2$), 
\be
R^2=1-\frac{\sum_{i}^{N}(y_{{\rm true},i}-y_{{\rm pred},i})^2}{\sum_{i}^{N}(y_{{\rm true},i}-\bar{y}_{{\rm true}})^2}
\ee
Thrid, the root mean squared error (RMSE), 
\be
\text{RMSE}=\sqrt{\frac{\sum_{i}^{N}(y_{{\rm true},i}-y_{{\rm pred},i})^2}{N}}.
\ee
Here, $y_{{\rm pred},i}$ and $y_{{\rm true},i}$ are the predicted value and the true (measured) value of the $i$-th data point, respectively. 
$N$ is the length of the data.
$\bar{y}_{\rm true}$ is the mean value of the true data. 
Throughout, we mainly use $R^2$ while showing the other two metrics for reference.


\section{Galaxy-halo connections} \label{sec:Results}

In this section, we present the models for predicting galaxy quantities with DM-halo quantities, as identified with the random-forest (RF) and symbolic-regression (SR) algorithms. 
We first test the performance of the regression models for global properties, specifically, stellar mass, star formation rate (SFR), and size, as a validation of the methods in \se{validation}, and then extend the analysis to disk properties in \se{diskpredictors}. 
For each relation, we first show the accuracy of our prediction, and then discuss the correlation between the target and the features.

\subsection{Predicting galaxy global properties with halo parameters}\label{sec:validation}

\subsubsection{Stellar mass}

{\bf Model accuracy:} 
Reassuringly, the stellar-mass-to-total-mass ratio is accurately predicted. 
RF attains high scores of $R^2=0.87$ using halo properties from the hydro runs, and $0.80$ using DMO halos, as shown in the upper panels of \Fig{validation}. 

{\bf Feature importance:}
The upper middle panel of \Fig{validation} presents the relative importance of the halo properties in predicting $M_\star/M_{\rm vir}$: halo mass $M_{\rm vir}$ dominates, followed by concentration $c$, formation redshift $z_{1/2}$, and accretion rate $\dot{M}_{\rm norm}$, consistent across both hydro and DMO predictions.
Some additional higher-order properties, including spin, shape parameters, and environment, may also contribute, with much smaller SHAP values.

{\bf Symbolic regression:}
Using the top 10 features, an empirical relation is obtained by training SR. 
It is worth noting that not all features remain in the final SR results after training. 
While the performance of SR model is not on par with the RF model, the resulting relations are simple and involve only four halo parameters, mass $M_{\rm vir,10}$, formation redshift $z_{1/2}$, concentration $c$, and short-to-long axis ratio $s$, consistent across both hydro and DMO versions. The explicit form for the DMO version is
\bad
\log \left(\frac{M_\star}{\Mv}\right) &=0.081z_{1/2}-1.375e^{-0.089M_{\rm vir,10}s}\\
&+0.042c-1.9.
\ead
This means that the scatter in the stellar-mass-halo-mass relation is linked to halo structure and formation history.
\Tab{relations1} and \Tab{relations1DMO} lists the functional forms of all the SR models involved in this study, for the hydro and DMO halos, respectively.
To further test the relations learned from SR, we show the stellar mass–halo mass relations at $z=0$ and $z=4$ in the upper right panel of \Fig{validation}. 
Excellent agreement exists between the SR predictions and the direct measurements from the TNG50 simulation. 
Remarkably, the relations based on the DMO halos are nearly identical to those based on the hydro halos.

\subsubsection{Star formation rate}
{\bf Model accuracy:} 
The middle panels of \Fig{validation} assess the models for the dimensionless star formation rate, SFR$t_{\rm H}/M_{\rm vir}$.
The model training and validation are both limited to the star-forming galaxies, i.e., with specific star formation rate higher than $10^{-2}$ Gyr$^{-1}$ \citep[e.g.][]{Wetzel13}. 
RF regression using the hydro and DMO halo measurements achieves $R^2$ values of 0.69 and 0.64, respectively. 

{\bf Feature importance:}
The most important parameter remains the virial mass. 
Other important halo properties, shared between the hydro and DMO versions, include formation redshift $z_{1/2}$, redshift $z$, spin $\lambda$, and the minor-to-major axis ratio $s$. 

{\bf Symbolic regression:}
SR gives an $R^2$ score that is slightly lower, and consistently picks up mass $M_{\rm vir,10}$, formation redshift $z_{1/2}$, and short-to-long axis ratio $s$ as the leading factors. 
The model based on the DMO halos is
\bad
\log\left(\frac{\text{SFR}}{\Mv/t_{\rm H}}\right)&=0.101z_{1/2}-1.296\\
&+e^{-0.146M_{\rm vir,10}s}\\
&\times(0.014M_{\rm vir,10}^{-2.337}-1.334).
\ead
Using the SR models, we compare the predicted star-forming main sequence against the simulation results in the right panel of \Fig{validation}. 
The SR predictions generally match the simulation result well, except for a slight excess at $M_\star \gtrsim 10^{10} \Msun$ at $z=0$.

\subsubsection{Galaxy size}

{\bf Model accuracy:} 
The lower panels of \Fig{validation} evaluate the accuracy of the regression models for the galaxy compactness, $r_{1/2}/\Rv$.  
RF regression using the hydro and DMO measurements achieve high $R^2$ scores of 0.76 and 0.62, respectively. 

\begin{figure*}
\centering
\includegraphics[width=0.87\textwidth]{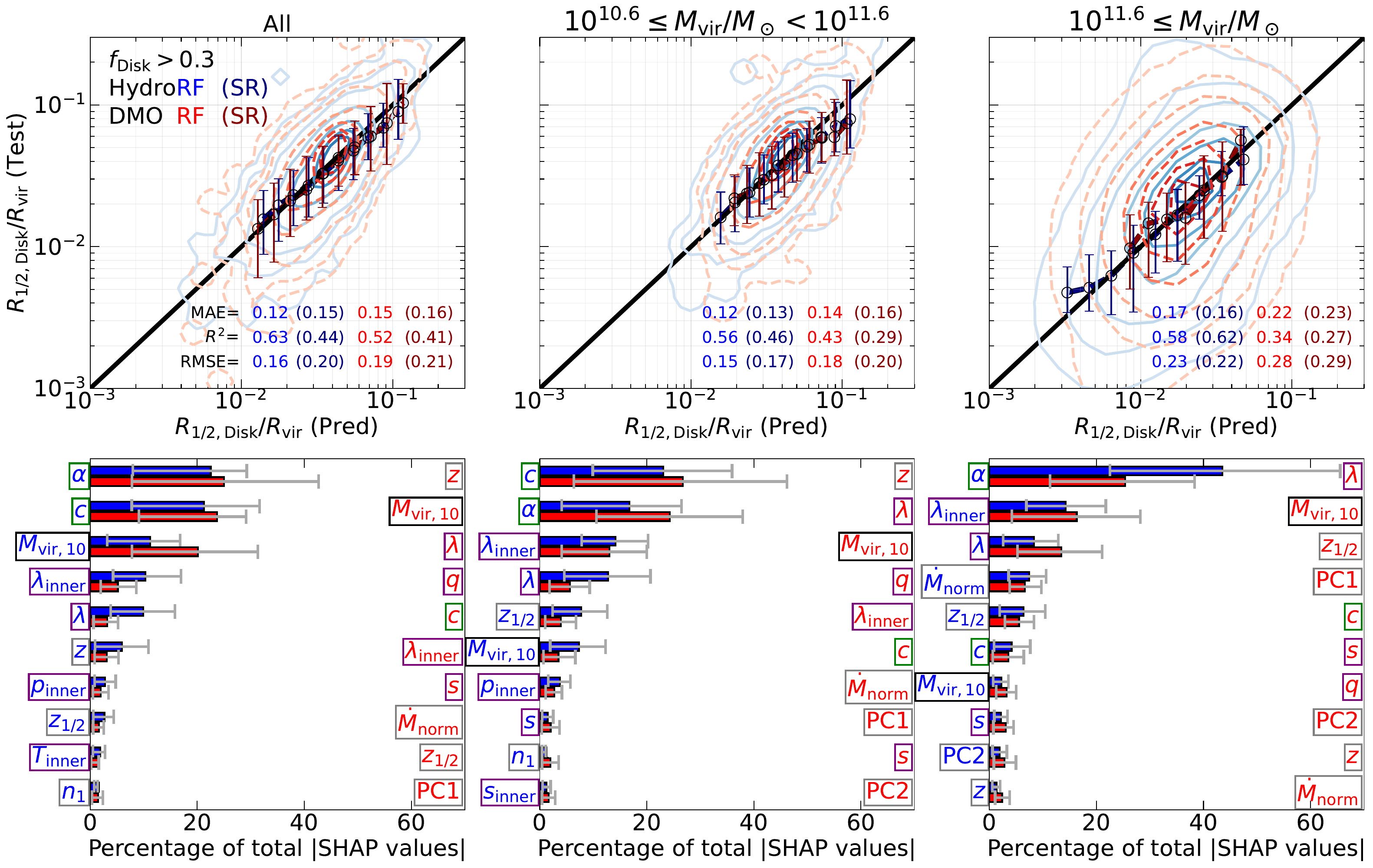}
    \caption{\textbf{Performance of regression models for disk compactness, $R_{\rm 1/2,Disk}/R_{\rm vir}$, and the most predictive halo properties} -- for galaxies with significant disk components ($f_{\rm Disk} > 0.3$). 
    Formatting follows \Fig{validation}. Columns correspond to different halo mass ranges: all ({\it left}), low-mass ($10^{10.6}\Msun \leq \Mv < 10^{11.6}\Msun$, {\it middle}), and high-mass ($M_{\rm vir} \geq 10^{11.6}\Msun$, {\it right}).} 
    \label{fig:DiskSize}
\end{figure*}

{\bf Feature importance:}
Unlike stellar mass or SFR, the most important halo property for galaxy compactness is not virial mass but redshift. 
In addition, the importance of virial mass is much smaller for the hydro halos compared to the DMO halos. 
For the hydro halos, the role of the Einasto shape index ($\alpha$) is significant, ranking the second, whereas it is not in the top 10 in the DMO case. 
This difference persists in predictions for disk structures (\se{diskpredictors}), and implies that $\alpha$ is significantly influenced by the galaxy rather than being a driver of galaxy size. 
Other commonly important properties across both hydro halos and DMO halos are concentration $c$, spin $\lambda$, and the inner-halo spin $\lambda_{\rm inner}$. 
A few environmental and assembly-history properties also weakly contribute to the prediction.

{\bf Symbolic regression:}
SR achieves performance scores that are slightly lower than RF, at $R^2=0.65$ (0.54) for the hydro (DMO) halos. 
The functional form for the DMO halos is
\bad
\log\left(\frac{r_{1/2}}{\Rv}\right)&=\log (\lambda+0.162)\log (M_{\rm vir,10}+c)+\log (z+s)\\
&+(0.643-0.201z^{0.463})(\log M_{\rm vir,10}+0.374)\\
&-1.474+\lambda_{\rm inner}+e^{-1.272M_{\rm vir,10}+0.642}.
\ead
The lower right panel of \Fig{validation} compares the size - stellar mass relations predicted by SR with those directly form the simulation. 
The regression models reproduce the data well at $z=0$.
They also match the low-mass end ($M_\star \lesssim 10^9$ M$_\odot$) at $z=4$ but exhibit a flatter high-mass end. 

\subsection{Predicting disk properties with halo properties}\label{sec:diskpredictors}

\subsubsection{Disk size}

{\bf Model accuracy:} 
\Fig{DiskSize} presents the regression models for disk compactness, $R_{\rm 1/2, disk}/\Rv$, where we have focused on galaxies with substantial disks, defined as $f_{\rm Disk}>0.3$. 
RF attains an $R^2$ core of 0.63 (0.52) with the hydro (DMO) halos. 
The regression results are consistently more accurate for the disks that are larger with respect to their host halos. 

\begin{figure*}	
\centering
\includegraphics[width=0.87\textwidth]{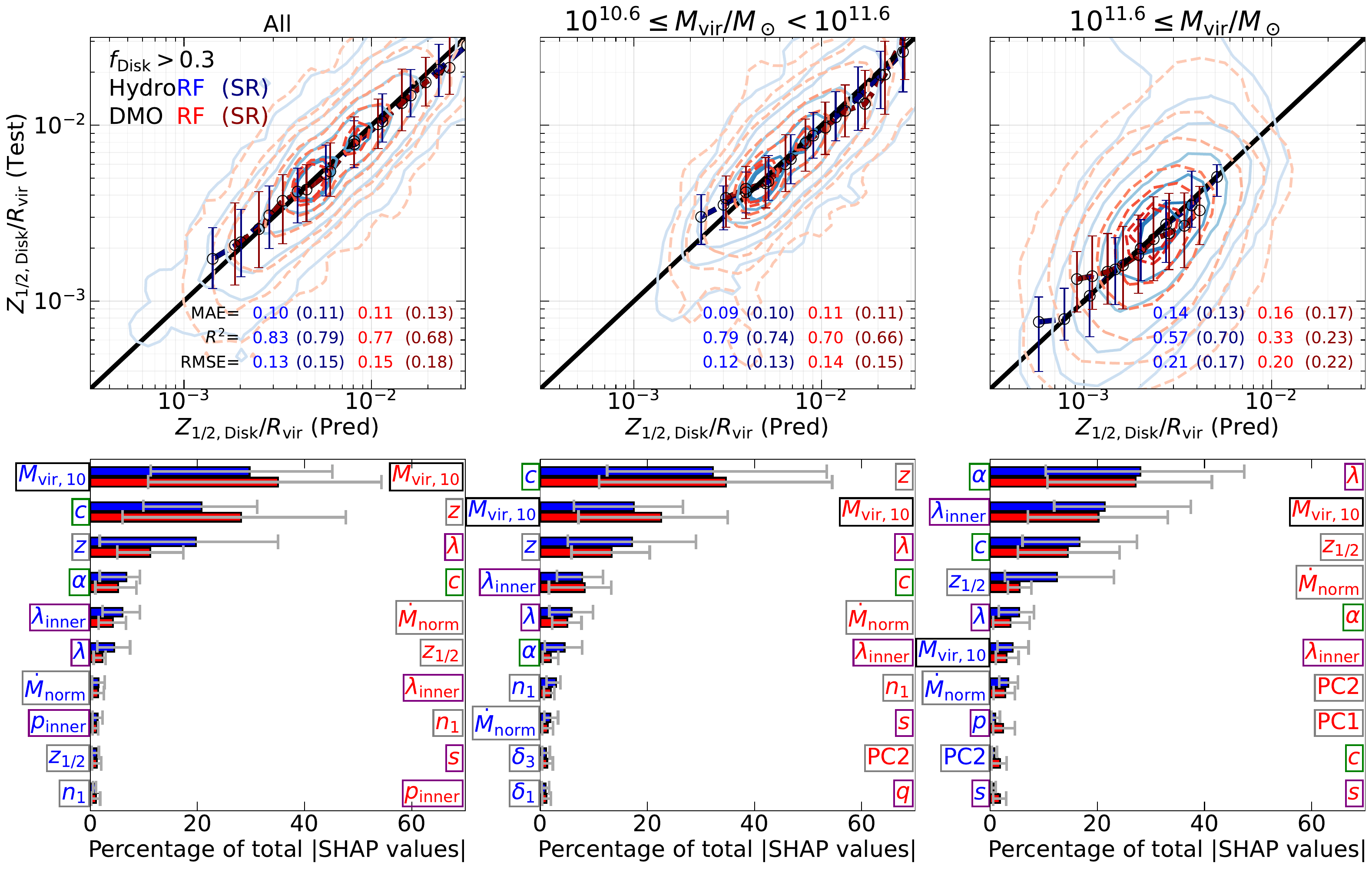}
    \caption{\textbf{Performance of regression models for disk thickness, $Z_{\rm 1/2,Disk}/R_{\rm vir}$,} for galaxies with significant disk components ($f_{\rm Disk} > 0.3$). 
    Formatting follows \Fig{validation}.
    Columns correspond to different halo mass ranges: all (left), low-mass ($10^{10.6} \leq M_{\rm vir}/M_\odot < 10^{11.6}$, middle), and high-mass ($M_{\rm vir}/M_\odot \geq 10^{11.6}$, right).
    }
    \label{fig:DiskHeight}
\end{figure*}

{\bf Feature importance:}
The importance rankings differ between the hydro and DMO cases. 
In the hydro case, the Einasto shape index $\alpha$ is the dominant factor for disk size, while it does not even appear in the top 10 in the DMO case. 
Some features are consistently influential for both hydro and DMO halos, including concentration $c$, spin $\lambda$, virial mass $M_{\rm vir}$, redshift $z$, and inner-halo spin $\lambda_{\rm inner}$. 
The most important halo properties align with those used in previous disk-size models \citep[e.g.][]{Mo98,Jiang19}.
Inner-halo parameters frequently emerge as important features when using the hydro halos, whereas global halo parameters tend to be more important when using the DMO halos. 
This difference arises mainly because the disk interacts via gravity with the surrounding DM that constitutes the inner halo, altering its structure, as will be further illustrated in \se{InnerHalo}.
Halo shape parameters are always among the top 10 features.
As discussed in \citet{Jiang25} and \citet{Liang25}, this is likely because rounder shapes are indicative of more relaxed halos in which extended disks can more easily develop.  
Environment and assembly-history parameters (e.g., $n_1$, PC1, PC2, $\dot M_{\rm norm}$) have limited impact. 

{\bf Symbolic regression:}
After training with the top 10 features, the SR algorithm reduces the inputs to only 4 parameters: inner-halo spin $\lambda_{\rm inner}$, total spin $\lambda$, Einasto shape $\alpha$, and concentration $c$, using the hydro halos. 
With the DMO halos, SR picks up 5 features -- virial mass $M_{\rm vir,10}$, redshift $z$, total spin $\lambda$, concentration $c$, and intermediate-to-long axis ratio $q$ -- and finds the following functional form
\bad\label{eq:disksize}
\log\left(\frac{R_{1/2,\rm Disk}}{\Rv}\right)&=\log M_{\rm vir,10}(\lambda^{0.251}-0.339)\\
&+z(\lambda+0.088)+0.343q-2.696\\
&+e^{-0.021c-0.007zM_{\rm vir,10}+0.045},
\ead
which is more complex and also less accurate than that based on the hydro halos.  
This again indicates that predicting disk structures using halo properties from $N$-body simulations is challenging, and suggests that the strong disk-halo connection in the hydro simulations is largely driven by baryonic processes.

{\bf Mass dependence:}
To better isolate secondary halo properties and assess potential mass dependence in the regression models, we divide the sample into two mass bins: $\Mv= 10^{10.6-11.6}\Msun$ and $\Mv>10^{11.6}\Msun$.  
The model behaviors for the two mass bins are illustrated in the middle and right columns of \Fig{DiskSize}.
Once halo mass is fixed, its relative importance naturally diminishes. 
In the low-mass bin, the leading features are concentration $c$, Einasto shape $\alpha$, inner-halo spin $\lambda_{\rm inner}$, and total $\lambda$ with the hydro halos; and redshift $z$, spin $\lambda$, inner-halo spin $\lambda_{\rm inner}$, axis ratio $q$, and concentration $c$ with the DMO halos.
The high-mass bin exhibits a stronger dependence on assembly-history parameters: accretion rate $\dot{M}_{\rm norm}$ and formation redshift $z_{1/2}$ rank fourth and fifth with the hydro measurements, while $z_{1/2}$ and PC1 rank third and fourth with the DMO measurements.
Paper I demonstrated that, at fixed mass, faster-accreting halos tend to host slightly larger disks. 
The present results align with this trend: because higher-mass halos generally grow more rapidly, accretion-related parameters exert a stronger influence on disk size in the high-mass sample.
Despite the mass-dependence of feature importance, SR basically identifies a consistent set of informative parameters. 
For the hydro halos, the features found by SR are inner-halo spin $\lambda_{\rm inner}$, spin $\lambda$, Einasto shape $\alpha$, and concentration $c$ (and also redshift $z$ for the high-mass bin). 
For the DMO halos, the useful features are virial mass $M_{\rm vir,10}$, redshift $z$, inner-halo spin $\lambda_{\rm inner}$, spin $\lambda$, concentration $c$, and axis ratio $q$ (and also accretion rate $\dot{M}_{\rm norm}$ for the high-mass bin).


\subsubsection{Disk scale height}\label{sec:DiskHeight}

{\bf Model accuracy:} 
The scale height $Z_{1/2}$, or thickness $Z_{1/2}/\Rv$, has not been linked to DM halo properties in previous studies. 
Interestingly as \Fig{DiskHeight} reveals, RF regression using halo properties attains accurate predictions for disk height, even more so than for disk compactness, with $R^{2}$ scores reaching 0.83 (0.77) with the hydro (DMO) halos. 

{\bf Feature importance:}
The halo properties most relevant for predicting disk thickness, as shown in the lower panels of \Fig{DiskHeight}, are nearly identical to those for disk compactness: 
they are Einasto shape $\alpha$, concentration $c$, virial mass $M_{\rm vir,10}$, and inner spin $\lambda_{\rm inner}$ for the hydro halos; and virial mass $M_{\rm vir,10}$, redshift $z$, spin $\lambda$, and concentration $c$ for the DMO halos. 
Environmental and assembly-history properties consistently rank higher in the DMO case than in the hydro case. 
For both cases, redshift $z$, formation redshift $z_{1/2}$, and accretion rate $\dot{M}_{\rm norm}$ are among the most influential top 10 features. 

{\bf Symbolic regression:}
The SR model for disk height is also more accurate than that for disk size.
The explicit functional form for the DMO halos is
\bad
\log\left(\frac{Z_{\rm 1/2,Disk}}{\Rv}\right)&=\lambda^{0.435}-0.461\log c-2.219\\
&+(p_{\rm inner}+z^{0.596})(0.534-0.237M_{\rm vir,10}^{0.224}).
\ead

{\bf Mass dependence:}
For the hydro halos, Einasto shape $\alpha$, concentration $c$, spin $\lambda$, inner spin $\lambda_{\rm inner}$, and redshift $z$ are always important predictors irrespective of the mass scale -- only the relative importance changes between the mass bins.
For the DMO measurements, however, the top features differ for different mass scales: redshift $z$, virial mass $M_{\rm vir,10}$, spin $\lambda$, concentration $c$, and accretion rate $\dot{M}_{\rm norm}$ are the most informative parameters for low-mass halos; whereas spin $\lambda$, virial mass $M_{\rm vir,10}$, formation redshift $z_{1/2}$, accretion rate $\dot{M}_{\rm norm}$, and Einasto shape $\alpha$ are the top factors for high-mass halos.

Overall, spin $\lambda$, inner spin $\lambda_{\rm inner}$, concentration $c$, mass $M_{\rm vir,10}$, and accretion rate $\dot{M}_{\rm norm}$ are the most important features for disk thickness.
Triaxial shape parameters are generally less influential for disk thickness than for disk compactness.

\subsubsection{Disk mass fraction}

\begin{figure}	
\centering
\includegraphics[width=0.75\columnwidth]{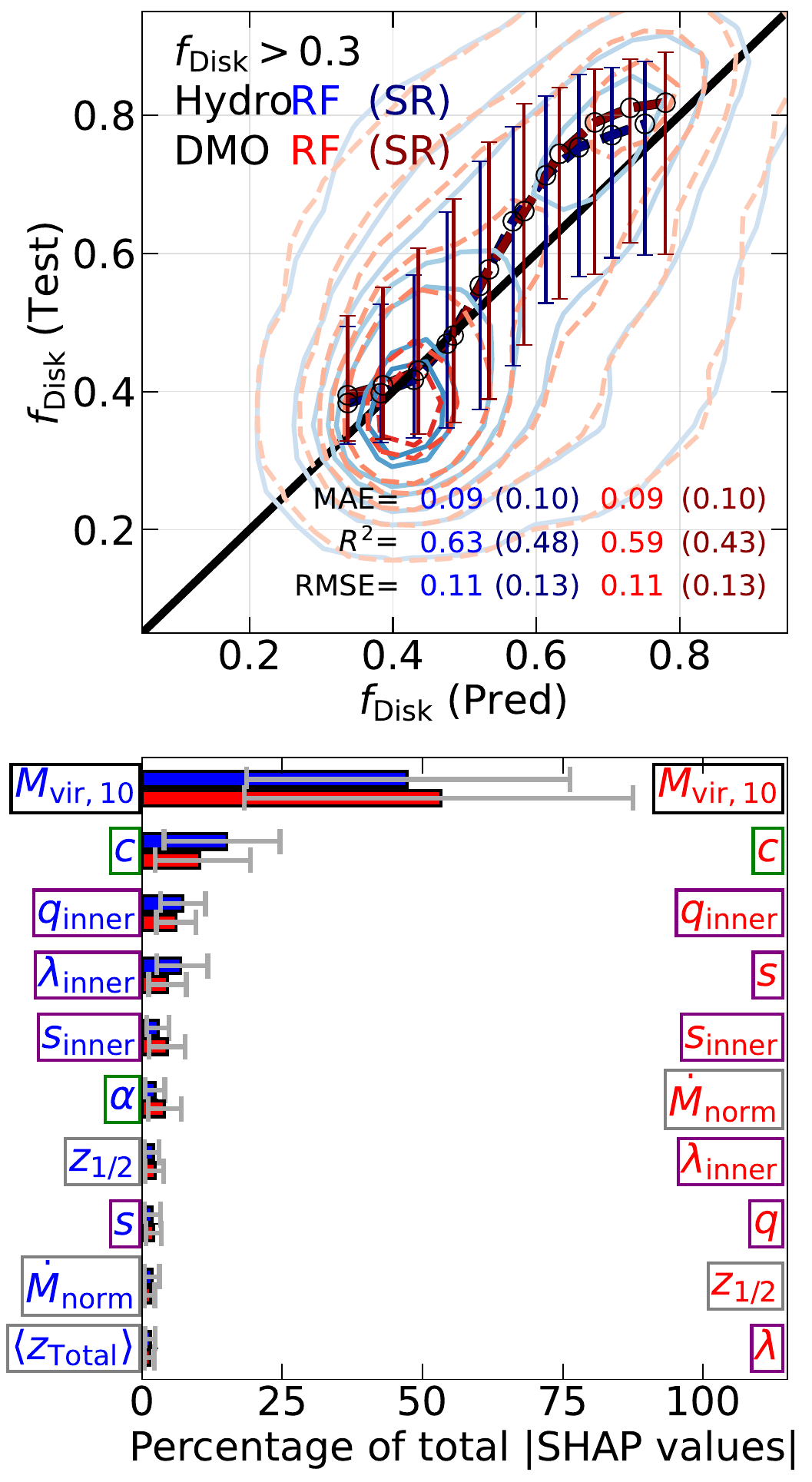}
    \caption{\textbf{Performance of regression models for disk mass fraction, $f_{\rm Disk}$,} for galaxies with significant disk components ($f_{\rm Disk} > 0.3$). 
    Formatting follows \Fig{validation}.}
    \label{fig:f_disk}
\end{figure}

{\bf Model accuracy:} 
For disk mass fractions, the sample is {\it not} divided into different halo mass bins, because the virial mass is by far the most dominant factor in regulating whether a significant disk can develop.
With the full sample, as shown in \Fig{f_disk}, RF attains an accuracy of $R^2\simeq 0.6$. 

{\bf Feature importance:}
Except for halo mass and concentration, the other halo properties have rather weak contributions. 
The relatively important features for both the hydro halos and the DMO halos are virial mass $M_{\rm vir,10}$, concentration $c$, inner shapes $q_{\rm inner}$ and $s_{\rm inner}$, as well as accretion rate $\dot{M}_{\rm norm}$. 

\begin{figure*}
\includegraphics[width=\textwidth]{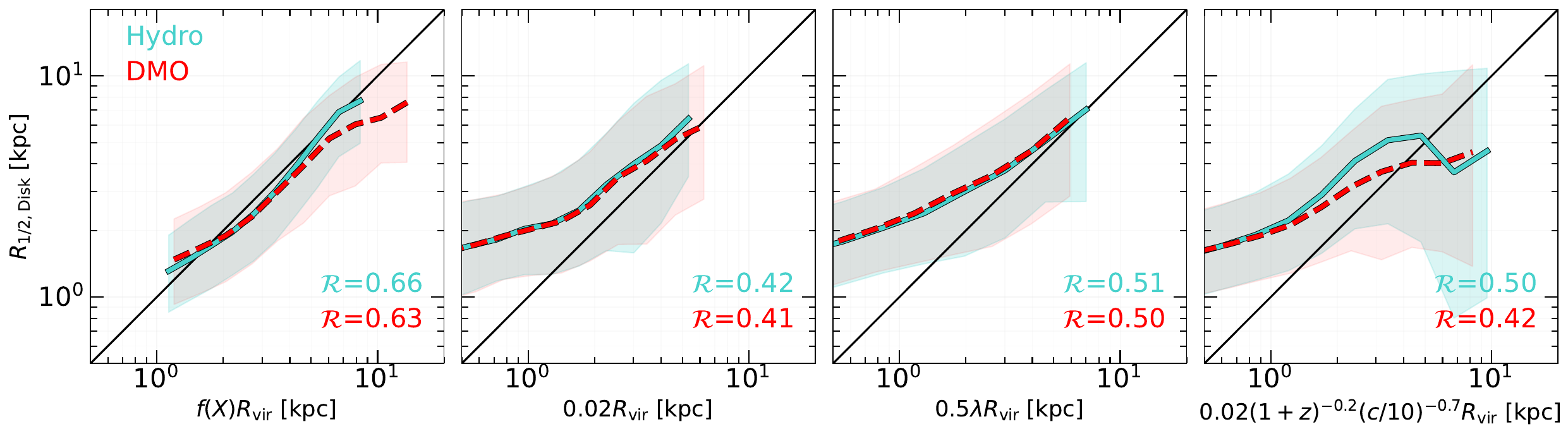}
    \caption{\textbf{Comparison of models relating disk size to halo properties.}
    Disk half-mass radii $R_{\rm 1/2,Disk}$ measured from the TNG50 hydro simulation are compared with predictions from several models.The first panel shows our SR-based model, written for simplicity as $R_{\rm 1/2,Disk} = f(X)R_{\rm vir}$, where $f(X)$ denotes the SR formula as a function of halo parameters $X$. The cyan and red curves correspond to the SR relations calibrated on hydro halos and DMO halos, respectively, with the DMO version given by \eq{disksize} and the hydro version by $f(X)=\lambda_{\rm inner}+(\lambda+c^{-0.125})(\log\alpha+3.010)-3.210$.
    The second panel shows the relation of \citet{Kravtsov13}, $R_{1/2}=0.02R_{\rm vir}$.The third panel shows a simplified version of the spin-based model of \citet{Mo98}, $R_{1/2}=0.5\lambda R_{\rm vir}$.The fourth panel shows the concentration-based empirical model of \citet{Jiang19}, $R_{1/2}=0.02(1+z)^{-0.2}(c/10)^{-0.7}R_{\rm vir}$.
    Median relations using hydro (DMO) halos are shown as solid cyan (dashed red) lines, with shaded regions indicating the 16–84th percentiles. Spearman correlation coefficients are listed in each panel, and the diagonal line marks the one-to-one relation.
    This analysis is limited to the galaxies with significant disks  ($f_{\rm Disk}>0.3$).
    \quad 
    Our SR models provide the highest correlation and lowest scatter, outperforming all previous prescriptions.}
    \label{fig:SizesComparisons}
\end{figure*}

{\bf Symbolic regression:}
The useful features picked up by SR are $c$ and $M_{\rm vir,10}$ for the hydro halo, and $M_{\rm vir,10}$ and $q_{\rm inner}$ for the DMO halos.
The explicit form of the SR model for DMO halos is
\bad
f_{\rm Disk}&=0.16q_{\rm inner}+c^{0.476}M_{\rm vir,10}(M_{\rm vir,10}+35.67)^{-1.31}\\
&+0.172.
\ead


\section{Discussion} \label{sec:Discussion}

In this section, we compare our new disk-size predictors with previous prescriptions, and interpret a few interesting aspects of the aforementioned disk-halo connections.

\subsection{New size predictors based on symbolic regression}

Simple analytic prescriptions have long been used to link galaxy size to halo properties.
\citet{Kravtsov13} derived a linear relation $R_{1/2} = 0.02\,R_{\rm vir}$ from abundance matching, applicable to both disks and spheroids.
\citet{Mo98} modeled thin exponential disks that inherit the halo’s specific angular momentum, yielding a size–spin scaling. 
Following \citet{Somerville18}, we adopt a much simplified form of the original \citeauthor{Mo98} model, $R_{1/2} = 0.5\,\lambda\,R_{\rm vir}$.
In contrast, \citet{Jiang19} found marginal correlation between halo spin and galaxy size in cosmological zoom-in simulations and proposed a concentration-based scaling,
$R_{1/2} = 0.02(1+z)^{-0.2}(c/10)^{-0.7}R_{\rm vir}$.

\Fig{SizesComparisons} contrasts previous analytic prescriptions with our SR-based models (see the $R_{1/2,\rm Disk}/\Rv$ entries in \Tab{relations1} and \Tab{relations1DMO}) by comparing their predicted disk sizes against the actual TNG50 measurements.
Our SR models deliver the highest fidelity, achieving Spearman coefficients $\mathcal R \gtrsim 0.6$—notably higher than the $\lesssim 0.5$ typical of earlier prescriptions.
Our models also exhibit reduced scatter and lie much closer to the 1:1 relation, whereas previous models systematically overpredict the sizes of compact disks.
With only a modest increase in input halo parameters compared to previous prescriptions, the SR relations achieve substantially higher accuracy, making them a practical choice for semi-analytic or empirical models built on $N$-body simulations.

\subsection{Understanding the relation between disk size and Einasto shape index}

\begin{figure*}	
\centering
\includegraphics[width=0.75\textwidth]{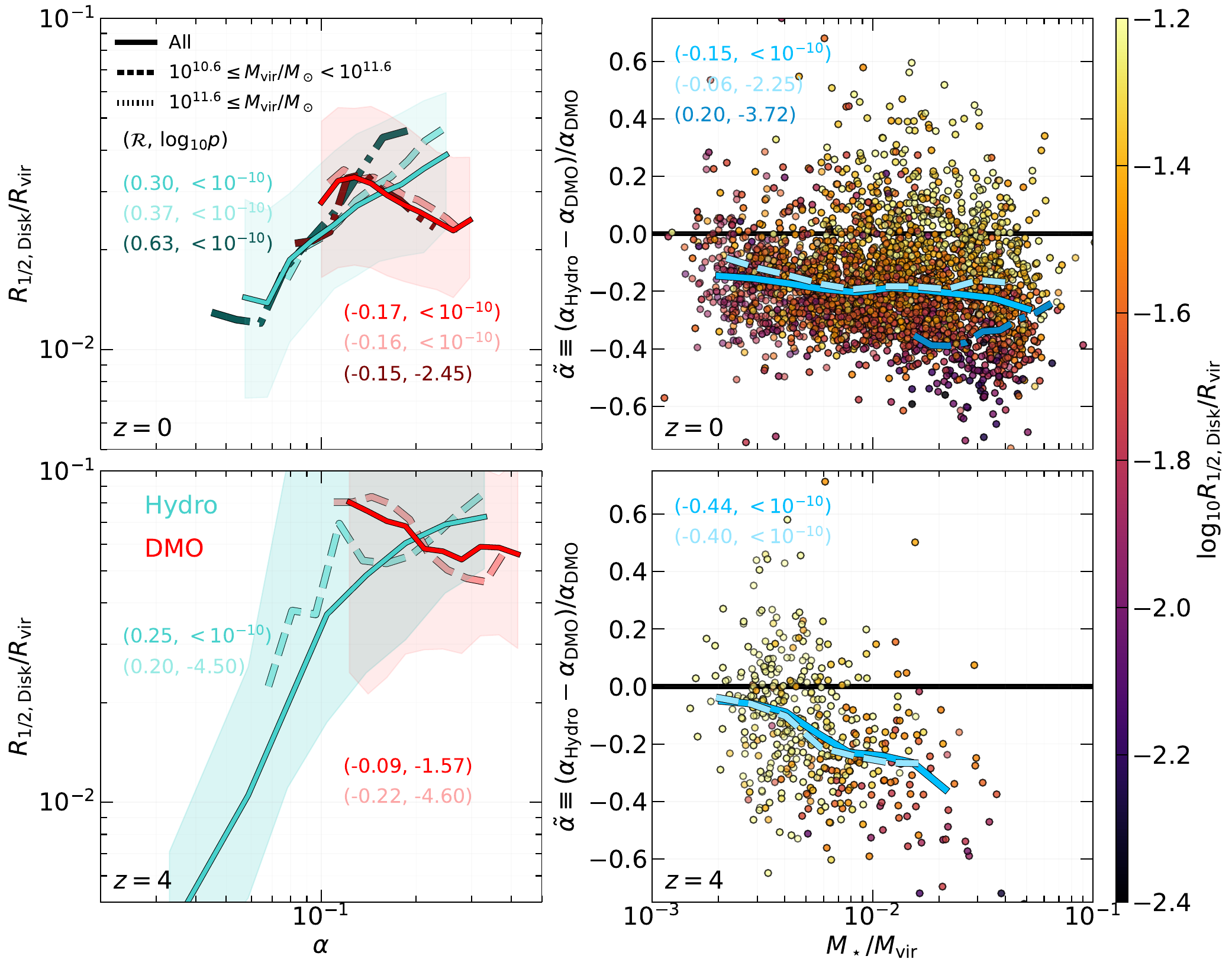}
    \caption{{\bf Halo structural response to disks.}
    The left panels show the relation between disk compactness, $R_{1/2,\rm Disk}/R_{\rm vir}$, and the Einasto shape index $\alpha$ of the host halo.
    The right panels show the ratio $\tilde{\alpha} \equiv (\alpha_{\rm Hydro}-\alpha_{\rm DMO})/\alpha_{\rm DMO}$ -- a measure of halo structural response to baryons -- as a function of the stellar-to-halo mass ratio, $M_\star/M_{\rm vir}$. 
    Here $\alpha_{\rm Hydro}$ is measured in the hydro run, and $\alpha_{\rm DMO}$ is that of the matched DMO counterpart.
    The top and bottom rows correspond to $z=0$ and $z=4$, respectively.
    In the left panels, hydro and DMO measurements are plotted in cyan and red, with dashed, solid, and dotted lines showing the low-mass, total, and high-mass samples (from light to dark colors). 
    Spearman correlation coefficients $\mathcal{R}$ and $\log p$ values are indicated, and shaded regions denote the 16–84th percentiles for the total sample.
    In the right panels, individual points are colored by disk compactness.
    This analysis is limited to the galaxies with significant disks  ($f_{\rm Disk}>0.3$).}
    \label{fig:slopeRd}
\end{figure*}

In \se{Results}, the SHAP analysis shows that the density–profile shape index, $\alpha$, is a key predictor of disk compactness, $R_{1/2,\rm Disk}/R_{\rm vir}$ -- but only when using halo properties measured from the hydrodynamic simulation. 
In the DMO case, $\alpha$ carries far less predictive power. 
This contrast suggests that the apparent link between disk size and $\alpha$ arises from baryonic effects, rather than from the primordial structure of DMO halos.
This is further explored in \Fig{slopeRd}, which shows a clear positive correlation between $R_{1/2,\rm Disk}/R_{\rm vir}$ and $\alpha$ in the hydro halos, but almost no correlation or weakly negative correlation in the DMO counterparts.

To directly test whether the $\alpha$–dependence of disk compactness is indeed driven by baryons, we define
\be
\tilde{\alpha} \equiv (\alpha_{\rm Hydro}-\alpha_{\rm DMO})/\alpha_{\rm DMO},
\ee
which quantifies the halo’s structural response to baryonic processes. Here, $\alpha_{\rm Hydro}$ is the Einasto index measured in the hydro halo, and $\alpha_{\rm DMO}$ is that of the matched DMO halo. By construction, $\tilde{\alpha}<0$ corresponds to halo contraction \citep{Blumenthal86, Gnedin04}, whereas $\tilde{\alpha}>0$ indicates expansion.
We also examine the stellar-to-halo mass ratio, $M_\star/M_{\rm vir}$, as a proxy for the strength of baryonic impact \citep{DC14, Freundlich20}. 
As shown in the right panels of \Fig{slopeRd}, most halos have $\tilde{\alpha}<0$, demonstrating that contraction is the dominant response in TNG50. 
Moreover, the mean value of $\tilde{\alpha}$ decreases with increasing $M_\star/M_{\rm vir}$, indicating that halos in more massive galaxies tend to contract more strongly.
At fixed stellar-to-halo mass ratio -- for example, around $M_\star/M_{\rm vir}\simeq 0.03$, where systems are comparably disk-dominated -- we find a clear trend that disk compactness governs the degree of halo contraction. The most extended disks (the yellow points in \Fig{slopeRd}) can even drive halo expansion rather than contraction.


\subsection{Redshift trend of disk compactness}

\begin{figure}	
\centering
\includegraphics[width=0.75\columnwidth]{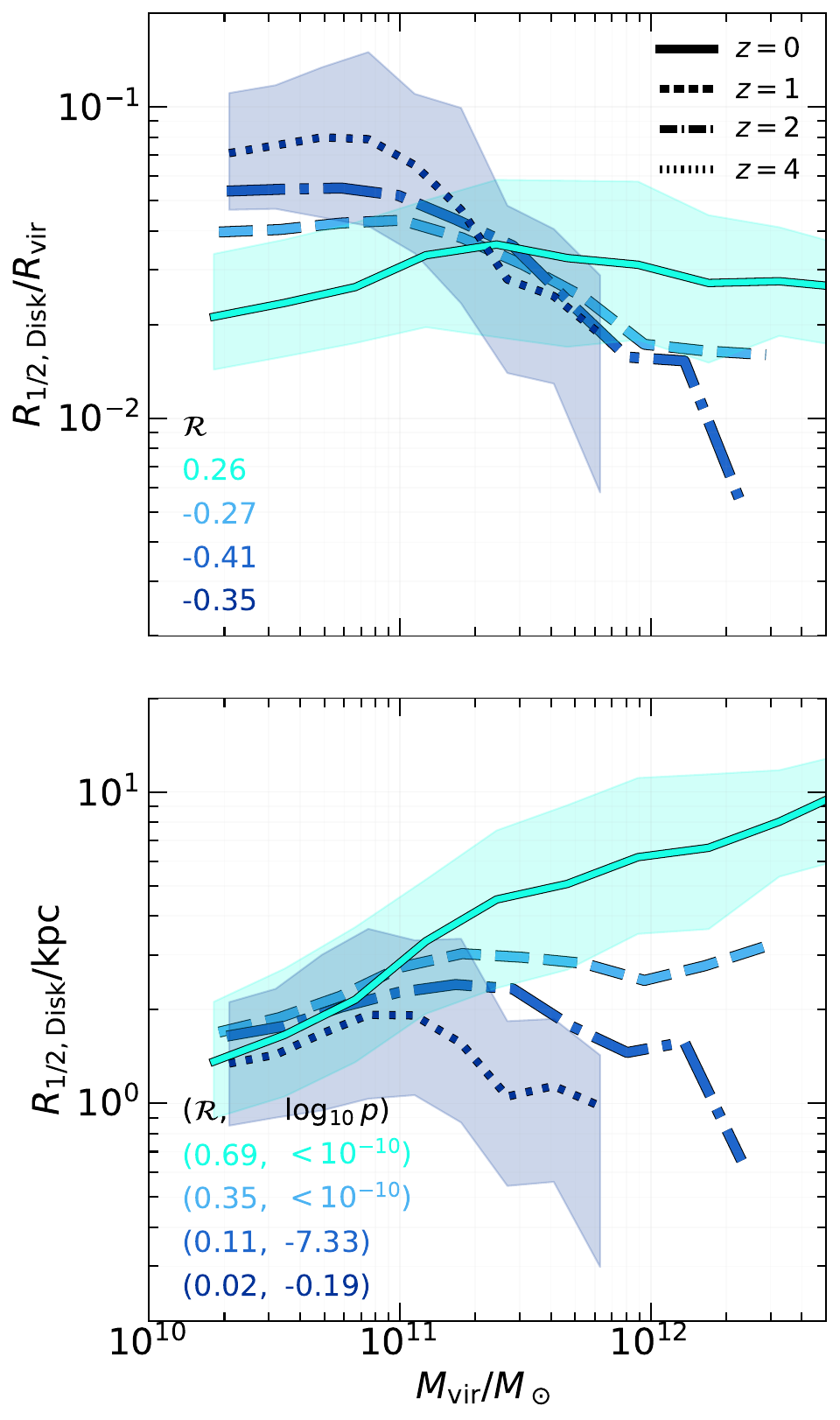}
    \caption{{\bf Disk compactness ($R_{1/2,\rm Disk}/R_{\rm vir}$) and physical size ($R_{1/2,\rm Disk}$) as functions of halo mass and redshift.} 
    Results at different redshifts are shown using distinct colors and line styles. 
    The Spearman correlation coefficient $\mathcal{R}$ and the corresponding $p$-value are reported. Shaded regions denote the 16–84th percentiles at $z=0$ and $z=4$.}
    \label{fig:RdRv}
\end{figure}

Here we discuss how disk size relative to halo size evolves with mass and redshift. 
As shown in the upper panel of \Fig{RdRv}, low-mass systems exhibit a strong redshift dependence: their $R_{\rm 1/2,Disk}/R_{\rm vir}$ increases significantly from $\sim 0.02$ at $z=0$ to $\sim 0.08$ at $z=4$. 
However, above a critical mass of $\Mv \simeq 2\times10^{11}M_\odot$, the trend reverses. 
Thus, the redshift evolution of disk compactness $R_{\rm 1/2, Disk}/R_{\rm vir}$ is mass dependent. 

A related trend is in the physical disk size $R_{1/2,\rm Disk}$, as shown in the lower panel of \Fig{RdRv}: it increases monotonically with halo mass at $z=0$, but at higher redshift it grows with halo mass below the critical mass and declines toward higher masses.

To understand these redshift trends, we first focus on the low-mass regime and ask what drives the strong evolution with redshift. A simple and intuitive picture emerges when one approximates the disk size as the radius at which gas first crosses the star-formation density threshold, and considers how the virial radius evolves at fixed halo mass. In the TNG model, star formation begins once the hydrogen number density exceeds $0.13\mathrm{cm}^{-3}$. Using a toy model with an NFW halo and a self-similar gas distribution containing the cosmic baryon fraction, we find that this star-forming radius remains roughly constant across redshift, primarily because typical halo concentrations decrease toward high $z$. By contrast, the virial radius, $\Rv \propto (M_{\rm vir}/\rho_{\rm crit})^{1/3}$, becomes much smaller at early times due to the rapid rise of the critical density. As a result, even if the physical disk size evolves only weakly, the ratio $R_{1/2,\rm Disk}/\Rv$ increases steeply toward high redshift. This picture is most applicable to low-mass halos, which predominantly grow via cold-mode accretion \citep{Dekel09}, where cooling is efficient and disk size closely tracks the extent of star-forming gas.

For more massive systems ($\Mv \gtrsim 2\times10^{11} M_\odot$), the processes that shape disk size differ markedly with redshift. At high $z$, these galaxies experience rapid mass assembly and frequent gas-rich mergers, which trigger centrally concentrated starbursts and produce compact stellar components \citep{Zolotov15,Lapiner23}. Although often described as star-forming nuggets with spheroidal morphologies, these systems can still retain substantial angular momentum inherited from merger orbits or inflowing filaments. Because our morphological decomposition identifies disks via the angular-momentum distribution of stellar particles, rapidly rotating nuggets are naturally classified as disks. This is not specific to our method -- most kinematics-based decomposition schemes behave similarly \citep[e.g.,][]{Du19,Zana22} -- reflecting the broader fact that high-$z$ disks and fast rotators are not cleanly separable. Consequently, the rotating stellar components in massive high-redshift halos tend to be compact.

At low redshift ($z\la 1$), by contrast, galaxy growth is more quiescent and proceeds inside-out \citep{Genel18}. Halos of a given mass are also more dynamically settled, providing favorable conditions for the formation of extended disks \citep{Jiang25}. This naturally explains the monotonic increase of disk size with halo mass at low redshift.

\subsection{Drivers of disk thickness: mass, size, and halo growth history}

As shown in \Fig{DiskHeight} and discussed in \se{DiskHeight}, halo properties exhibit unexpectedly strong predictive power for disk thickness, stronger, in fact, than for disk compactness.
Here we explore clues to the possible physical origin of this behavior.

\begin{figure*}	
\centering
    \includegraphics[width=0.75\textwidth]{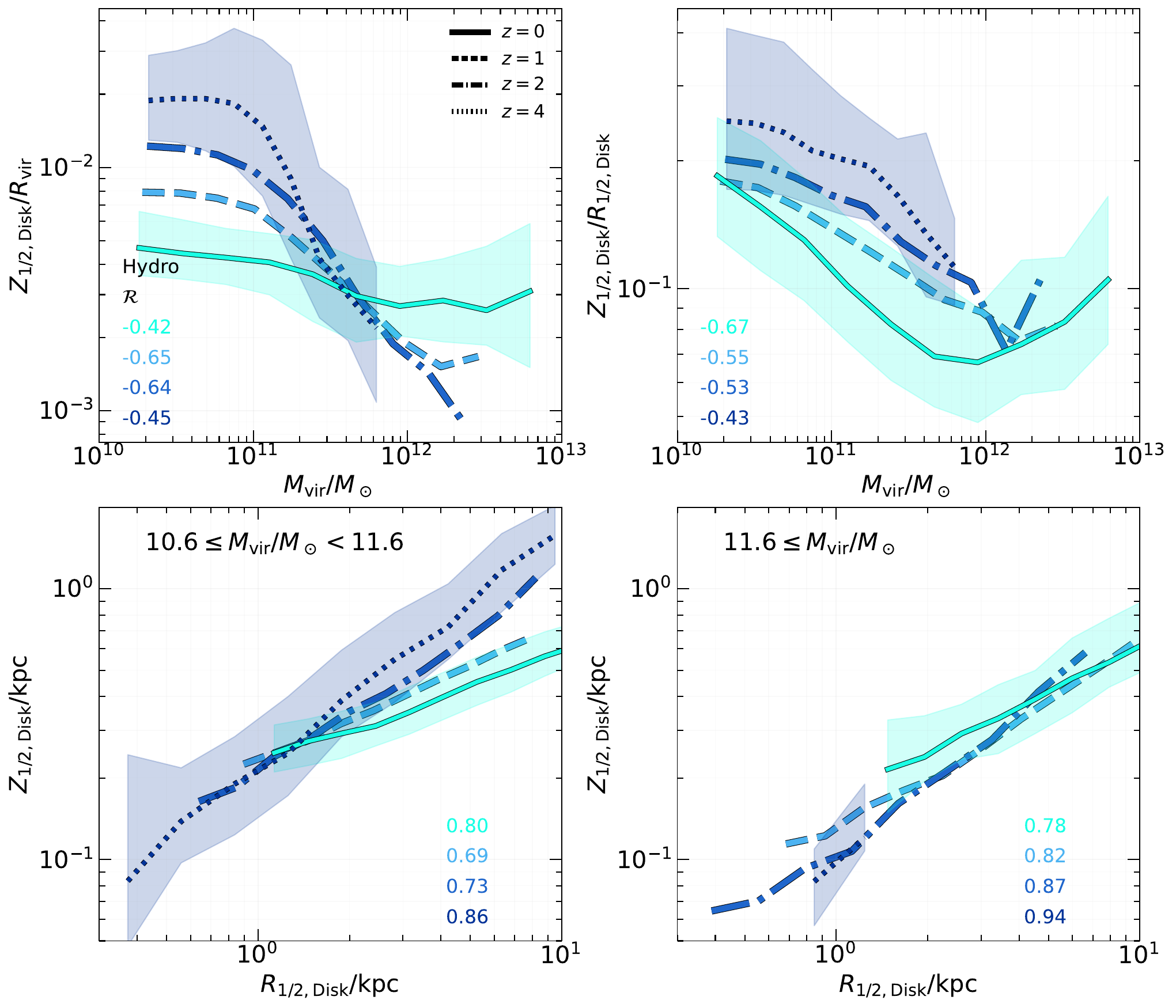}
    \caption{
    {\it Upper left}: Disk thickness as a function of halo mass and redshift. 
    {\it Upper right}: Disk flattening, quantified by the ratio between height and size, $Z_{\rm 1/2,Disk}/R_{\rm 1/2,Disk}$, as a function of halo mass and redshift.
    {\it Middle} and {\it right}: Disk height $Z_{\rm 1/2,Disk}$ versus size $R_{\rm 1/2,Disk}$ for two halo-mass bins, $\Mv=10^{10.6-11.6}\Msun$ and $\Mv > 10^{11.6}\Msun$.
    Line styles and colors indicate different redshifts. 
    Spearman correlation coefficients $\mathcal R$ are listed in each panel.
    Shaded regions show the 16–84\% ranges at $z=0$ and $z=4$.
    \quad
    Milky-Way mass halos host the thinnest and most flattened disks. 
    At fixed mass (focusing on the dwarf regime of $\Mv\le10^{11.3}\Msun$), disks thicken towards higher redshift.
    Disk size and height are strongly correlated, implying that the halo factors that drive disk size also regulate disk height. 
    }
    \label{fig:DiskSizeHeightMvir}
\end{figure*}

We begin by examining how disk thickness depends on halo mass and redshift.
As shown in \Fig{DiskSizeHeightMvir}, disks become progressively thinner (relative to the virial radius) as halo mass increases from the dwarf regime up to the Milky Way scale, with the trend especially pronounced at high redshift.

The disk flattening ratio, $Z_{\rm 1/2,Disk}/R_{\rm 1/2,Disk}$, reaches a minimum of $\sim 0.1$ near $\Mv\sim10^{12}\Msun$ and rises at both lower and higher masses.
At any fixed mass, disks are systematically thicker at earlier cosmic times.
This potentially reflects enhanced turbulence and elevated vertical velocity dispersion in high-redshift galaxies, consistent with vertical Jeans equilibrium, where $Z_{1/2}\propto \sigma_z^2 / (\pi G \Sigma)$.

These strong mass and redshift dependencies agree with the SHAP rankings in \Fig{DiskHeight}, which consistently identify halo mass and redshift as the dominant predictors of disk thickness.

A second key finding is the tight correlation between disk height and disk size.
As shown in the middle and right panels of \Fig{DiskSizeHeightMvir}, disk height increases almost linearly with disk size at fixed halo mass.
This implies that many halo properties might influence disk thickness primarily through their regulation of disk size—so the factors that set disk extent also tend to set disk thickness.

\begin{figure*}	
\centering
\includegraphics[width=0.75\textwidth]{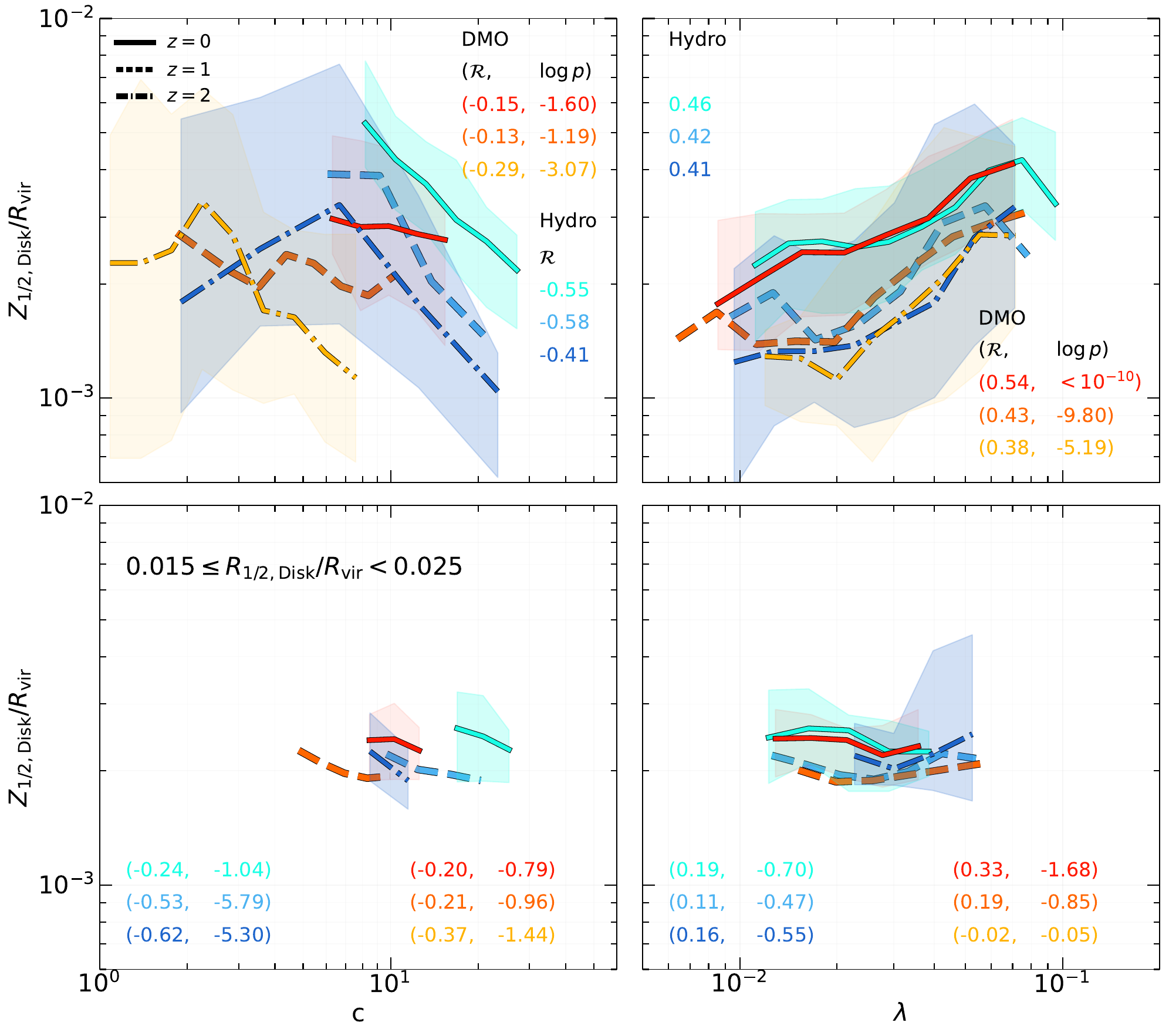}
    \caption{Relations between disk thickness and halo concentration ({\it left}) and spin ({\it right}), evaluated at fixed halo mass ($10^{11.7}<\! M_{\rm vir}/M_\odot \!<\! 10^{12.3}$, basically the Milky-Way scale).
    The lower row further restricts the sample to a narrow range of disk compactness ($0.015 \le R_{\rm 1/2,Disk}/\Rv < 0.025$).
    Line styles denote different redshifts, while warm and cool colors distinguish results from the hydro and DMO halos.
    Spearman correlation coefficients $\mathcal R$ and the corresponding $p$-values are quoted.
    Shaded regions show the 16–84\% ranges at $z=0$ and $z=2$.}
    \label{fig:Zdisk_c_spin}
\end{figure*}

We then examine in \Fig{Zdisk_c_spin} the correlations of thickness with halo structural parameters, spin and concentration, which are the most important factors after mass and redshift according to the SHAP scores. 
Here we fix the halo mass to the Milky-Way scale ($\Mv=10^{11.7-12.3}\Msun$) to better focus on the trend with the structural parameters.
Disk thickness increases with spin for both hydro halos and DMO halos, while a dependence on concentration appears only in the hydro case.
Interestingly, once disk compactness is fixed, as for the lower panels of \Fig{Zdisk_c_spin}, these correlations largely disappear. 
This may suggests that spin and concentration affect thickness indirectly -- through their impact on disk size and compactness -- rather than by directly regulating the vertical velocity dispersion.

\begin{figure*}	
\centering
\includegraphics[width=0.75\textwidth]{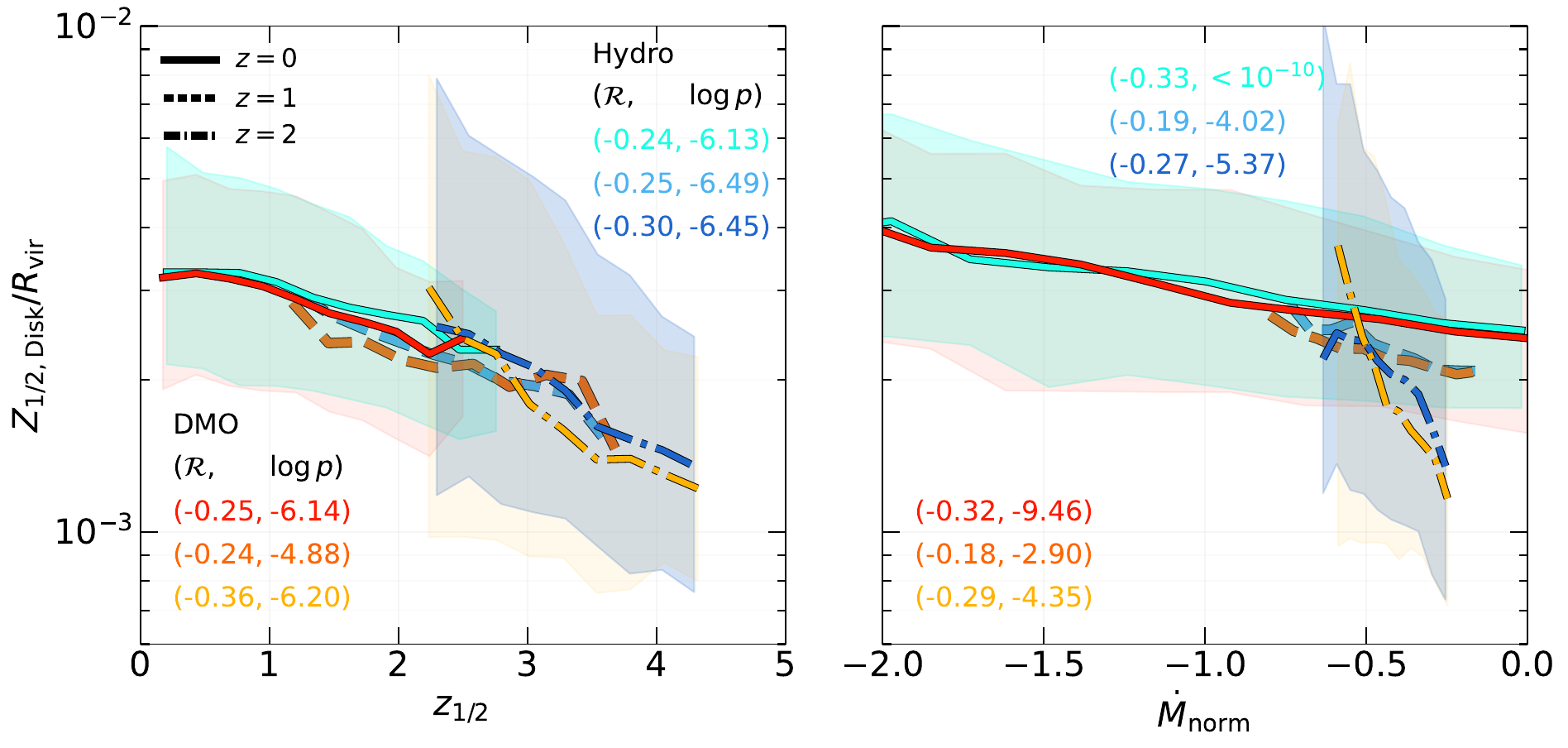}
    \caption{
    Relations between disk thickness $Z_{1/2,{\rm Disk}}/R_{\rm vir}$ and halo assembly-history parameters: formation redshift $z_{1/2}$ ({\it left}), and accretion rate $\dot{M}_{\rm acc}$ ({\it right}). Formatting follows \Fig{Zdisk_c_spin}.}
    \label{fig:Zd_Zh_Macc}
\end{figure*}

Finally, we examine disk thickness versus halo assembly–history parameters in \Fig{Zd_Zh_Macc}.
At fixed halo mass, lower formation redshift $z_{1/2}$ and lower normalized accretion rate $\dot{M}_{\rm norm}$ -- corresponding to more recent or more rapid mass growth -- produce systematically thicker disks at all redshifts.
This is a natural consequence of dynamical heating from mergers and accretion.
Taken together, these results might indicate that active mass assembly is the primary driver of disk thickness, whereas the apparent correlations with halo spin and concentration arise indirectly through their shared impact on disk compactness.
Although a full explanation of how halo structure governs disk compactness and why disk size and height correlate so tightly lie beyond the scope of this work, these results provide a clear clue as to why halo properties predict disk thickness better than disk size.
The additional predictive power possibly comes from mass-assembly parameters, which exert a stronger influence on disk thickness than on disk compactness.

\subsection{Do inner halo properties improve disk–halo modeling?}\label{sec:InnerHalo}

\begin{figure*}	
\centering
\includegraphics[width=0.75\textwidth]{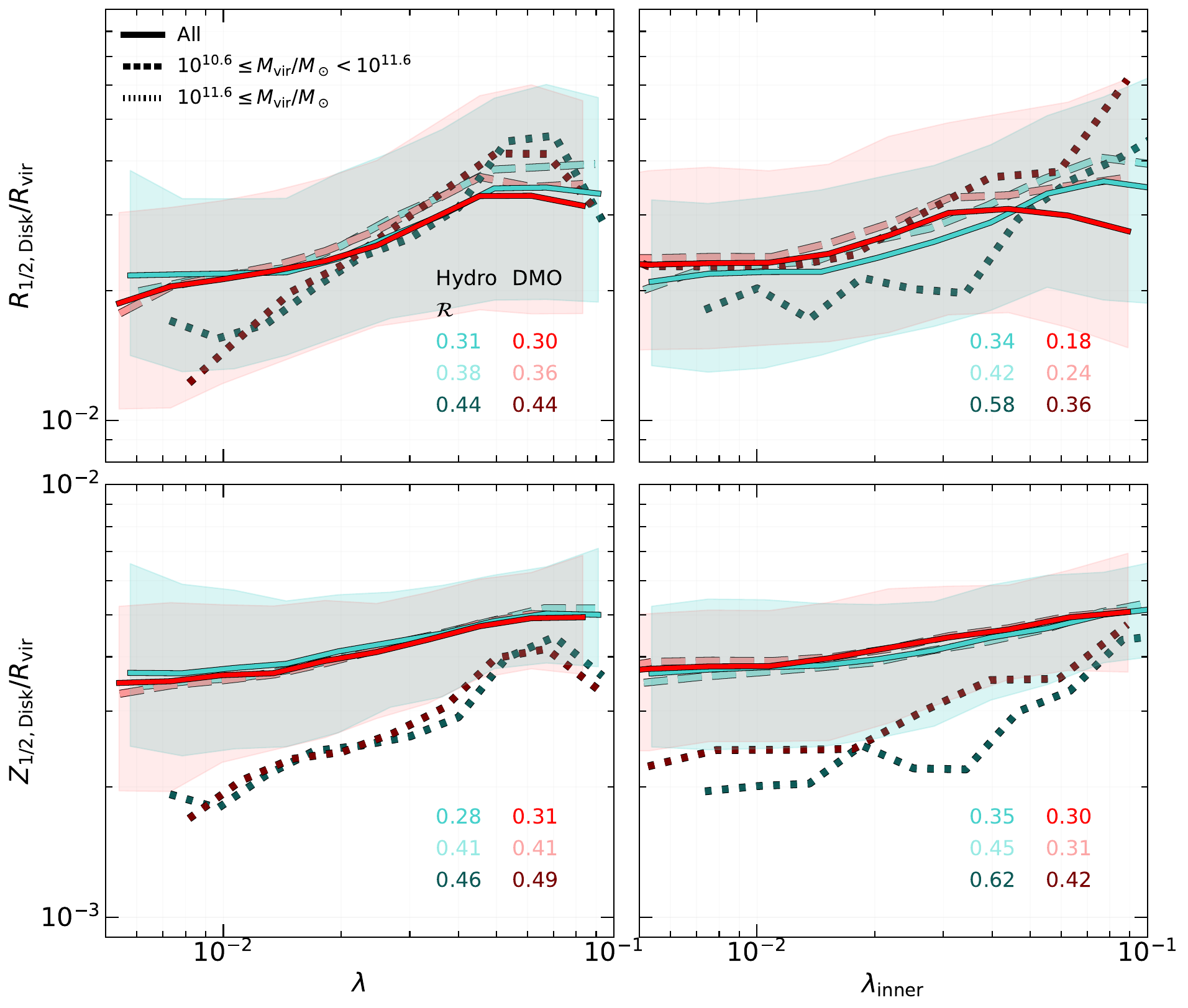}
    \caption{Disk compactness $R_{\rm 1/2, Disk}/\Rv$ ({\it top}) and thickness $Z_{\rm 1/2, Disk}/\Rv$ ({\it bottom}) as functions of the global halo spin $\lambda$ ({\it left}) and inner-halo spin $\lambda_{\rm inner}$ ({\it right}). 
    Formatting follows \Fig{slopeRd}. 
    Spearman correlation coefficients $\mathcal{R}$ are indicated.
    \quad Inner-halo spin exhibits a noticeably stronger correlation with disk properties in the hydro halos, especially in the high-mass regime, whereas for the DMO halos the global-spin correlations are marginally stronger.
    }
    \label{fig:RdRvInnerHalo}
\end{figure*}

As shown earlier in \se{Results}, inner-halo parameters emerge as highly important predictors of disk structure—but only when measured from the hydrodynamical simulation.
To clarify this difference, \Fig{RdRvInnerHalo} compares the predictive power of the global spin $\lambda$ and the inner-halo spin $\lambda_{\rm inner}$ for both disk compactness and disk thickness.

Two key trends appear. 
First, for the hydro halos, $\lambda_{\rm inner}$ correlates more strongly with both disk size and thickness than the global spin, with the effect being particularly pronounced in the high-mass sample ($\Mv>10^{11.6}\Msun$).
Second, for the matched DMO halos, the situation reverses: correlations with the global spin are marginally stronger than those with $\lambda_{\rm inner}$.

This contrast highlights the role of baryonic physics.
In the hydro run, the disk exchanges angular momentum with the surrounding dark matter and reshapes the inner halo, creating a tighter disk -- inner halo coupling.
In the DMO run, no such interaction occurs, and the inner-halo quantities do not carry additional predictive power beyond the global ones.

Thus, for semi-analytic models or empirical galaxy–halo frameworks built on $N$-body–only halos, global halo properties are not only sufficient but may in fact be more robust predictors than inner-halo quantities.

\subsection{Generalizability beyond TNG physics}

Our results are based on the TNG simulations. Therefore, the feature importance identified by the RF and SHAP analyses, the empirical relations learned by SR, and the correlations are specific to TNG physics rather than universally applicable. 
For example, halo spin parameters emerge as key predictors of disk structure in our analysis. 
Similarly strong correlations with spin are also reported in Paper I and in \cite{Liao19} using the Auriga simulation \citep{Grand17}. 
However, other simulations may show weaker spin dependence. \cite{Rohr22} based on FIREbox \citep{Feldmann23}, \cite{Jiang19} based on NIHAO \citep{Wang15} and VELA \citep{Ceverino14,Zolotov15} find that halo spin is not strongly linked to galaxy size. 
\cite{Yang23} demonstrated that correlations between galaxy compactness and halo spin are significantly stronger in TNG than in EAGLE. 
We speculate that these differences arise from more efficient baryonic cycles, and will follow up in a future study.

Therefore, applying the same analytical framework to those simulations may not highlight the spin parameter as an important factor, and the SR relations may not perform well using halo properties from those simulations. 

To test whether the regression relations remain robust when the subgrid physics prescriptions are varied requires applying the models to a different simulation suite. 
While such a comparison is beyond the scope of the current work, we can perform a related internal test within the TNG sample by examining whether the models hold for subsamples selected by specific star formation rate (sSFR).
A primary effect of modifying subgrid prescriptions is to alter the global star-formation activity of galaxies. Therefore, if the model performs well for subsamples split by sSFR, this would suggest that it is relatively insensitive to such recipe variations.

\begin{figure}	
\centering
\includegraphics[width=0.75\columnwidth]{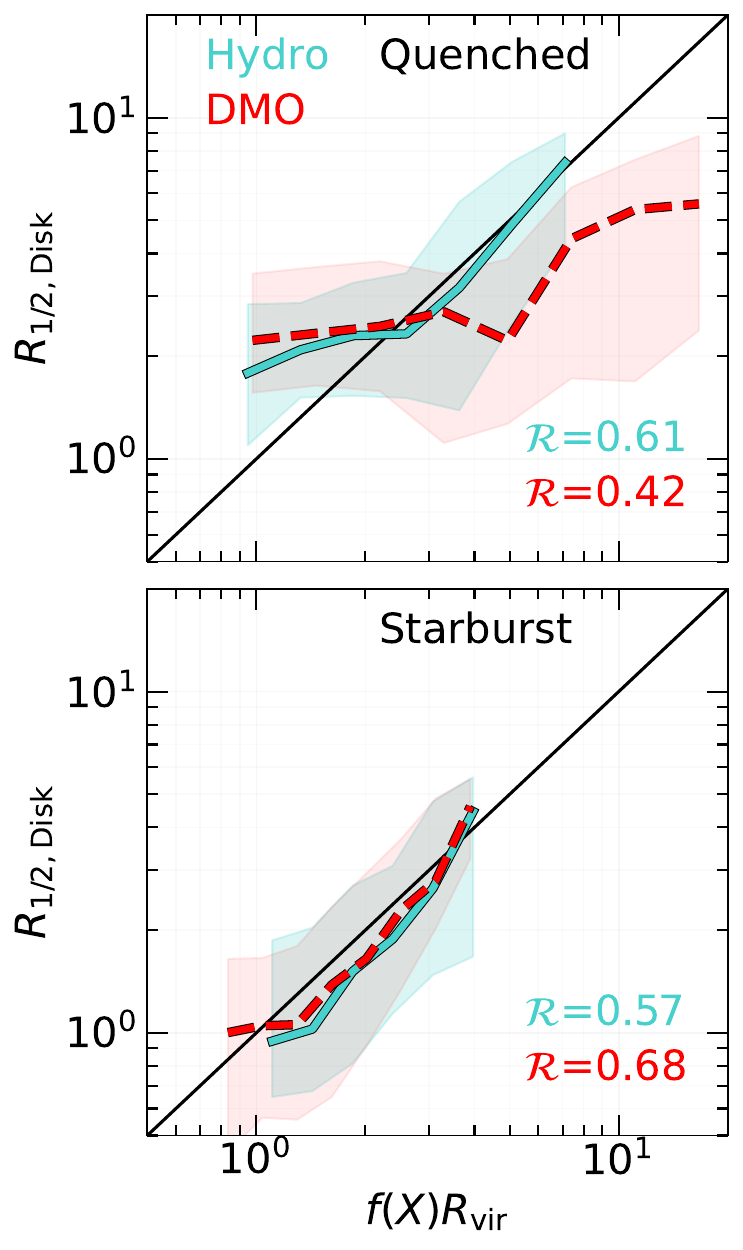}
    \caption{{\bf Evaluation of the applicability of the SR size predictors to galaxy subsamples.}
    Similar to \Fig{SizesComparisons}, but the SR relations are applied only to quenched galaxies ($\mathrm{SFR}/M_\star \leq 10^{-11}\ \mathrm{yr}^{-1}$) or to starburst galaxies ($\mathrm{SFR}/M_\star \geq 10^{-8.5}\ \mathrm{yr}^{-1}$).
    \quad
    The Hydro SR relations remain highly effective even for quenched galaxies (and are thus likely insensitive to subgrid physics, which primarily affects the star-formation status), whereas the weaker DMO performance indicates that baryonic processes associated with quenching substantially modify halo properties in ways not captured by DMO size predictor.
    }
    \label{fig:subsample}
\end{figure}

In \Fig{subsample}, we evaluate our size predictors for quiescent galaxies ($\mathrm{SFR}/M_\star \leq 10^{-11}\,\mathrm{yr}^{-1}$) and for starburst galaxies ($\mathrm{SFR}/M_\star \geq 10^{-8.5}\,\mathrm{yr}^{-1}$). 
These two populations lie in the $\sim 2$\% tails of the sSFR distribution in our dataset. 
The SR relations continue to perform well for starburst galaxies in both the hydro and DMO runs, achieving correlations of $\mathcal{R} \gtrsim 0.6$. 
For quenched galaxies, the hydro-based model also attains $\mathcal{R} = 0.61$, whereas the DMO version performs less well ($\mathcal{R} = 0.42$).

These results indicate that the hydro SR relations are likely not very sensitive to details of the subgrid physics and remain effective predictors of disk sizes even for quenched subsamples. 
In contrast, the weak performance of the DMO model implies that baryonic processes associated with quenching might substantially alter halo properties -- such information is missing in the DMO halos.

\subsection{Resolution effects on disk structures}
To test whether disk structures (e.g. size and height) are well resolved, we examine the distributions of disk size and disk height and compare them with the collisionless softening length (0.288 ckpc). 
For disk size, the 16th percentile is above the softening length for all samples at all redshifts. 
For disk height, the 16th percentile for total samples at $z=0$ and for MW-mass systems at $z=2$ and $z=4$ is comparable to the softening length, indicating marginally sufficient resolution for some cases. 
We experimented with restricting our analysis to well-resolved subsamples, and find similarly high prediction accuracy for both our RF and SR models, suggesting that the ML models might capture underlying physical correlations rather than being sensitive to numerical resolution effects. However, applying our pipeline to higher resolution simulations might still be needed since some cases are marginally resolved.

\subsection{Sample completeness at low masses}

Following Paper I, we require a minimum of 1000 stellar and 1000 DM particles. 
The threshold on the number of stellar particles ensures robust morphological decomposition but is the more restrictive condition of the two, since galaxies with 1,000 stellar particles typically reside in halos well above 1000 DM particles.
For this reason, our sample and the measurements for halo quantities becomes incomplete at the low-mass end. 
To qualitatively describe this bias: galaxies that are strongly DM-dominated are missing from our analysis at the low-mass end.

\subsection{The selection of best-fitting functions}
The SR requires a metric to select functions at each iteration, with MSE-like losses (e.g. \Eqnb{NMSE}, \Eqnb{NQL}) being the simplest choice. However, since such losses typically decrease with increasing complexity, model selection must balance accuracy and simplicity, often via heuristic penalties or Pareto fronts. These approaches may under-penalize complexity. Information criteria such as Akaike (AIC), Bayesian (BIC), and Deviance (DIC) provide an alternative by penalizing parameter count, but they do not capture functional complexity. A more principled approach is the minimum description length (MDL) principle \citep[e.g.][]{Grunwald19}, which jointly quantifies accuracy and complexity from an information-theoretic perspective and has been applied in astrophysical SR studies \citep[e.g.][]{Bartlett24,Harry25}. Here we adopt MSE-like losses for simplicity and good empirical performance, while encouraging future work to explore metrics that better penalize functional complexity.


\section{Conclusion} \label{sec:Conclusion}

This work presents a comprehensive investigation of the connections between galaxy properties and host dark-matter halo properties in the TNG50 cosmological simulation, focusing on predicting disk structure using halo properties. 
We measure 37 halo properties -- capturing density structure, angular momentum, shape, formation history, and environment -- from both the full-physics (hydro) simulations and dark-matter-only (DMO) runs. 
Using the morphological decomposition method developed in Paper I, we obtain disk properties (size, scale height, and mass fraction), and also analyze global galaxy quantities including stellar mass, SFR, and size. Random Forest (RF) and Symbolic Regression (SR) are then used to assess predictability, determine the most informative halo features, and derive empirical relations. Our main conclusions are the following.

Regression models predict global galaxy properties with very high accuracy.
Simple SR expressions successfully reproduce the stellar–halo mass relation, the star-forming main sequence, and the size–mass relations across redshifts.
These relations offer valuable prescriptions for empirical modeling and demonstrate the promise of machine-learning-based approaches for predicting more complex disk structures from halo information.

Regarding the predictability of disk structure, we find
\begin{itemize}[leftmargin=*]
\item Disk properties are predictable with good accuracy.
SR performs slightly worse but still captures the major trends.
\item Disk height is easier to predict than disk size, and predictions are more accurate for dwarfs ($\Mv<10^{11.6}\Msun$) than for more massive systems.
\item Models based on hydro halos consistently outperform DMO-based ones, and the DMO-based SR relations are more complex. 
\end{itemize}

Different aspects of disk structure depend on different halo parameters.
\begin{itemize}[leftmargin=*]
\item  Disk compactness, defined as the half-mass size in units of the halo virial radius, is strongly linked to the Einasto shape index $\alpha$, concentration $c$, inner-halo spin $\lambda_{\rm inner}$, total spin $\lambda$, as well as the formation redshift $z_{1/2}$, using hydro halos. Using the DMO halos, which are typically what semi-analytical models work with, the most predictive halo parameters are spin $\lambda$, halo mass $\Mv$, redshift $z$, concentration $c$, and the short-to-long axis ratio $s$. 
\item Disk thickness, defined as the half-mass height in units of the halo virial radius, can be accurately predicted by halo mass $\Mv$, concentration $c$, inner-halo spin $\lambda_{\rm inner}$, total spin $\lambda$, and accretion rate $\dot{M}_{\rm norm}$, together with the Einasto shape index $\alpha$ or redshift $z$ for the hydro halos and DMO halos, respectively. 
\item Disk mass fraction is largely determined by halo mass, with secondary dependence on concentration and inner-halo shape. 
\end{itemize}
Across all targets, RF and SR consistently identify concentration, spin, mass, and assembly history as the most predictive features. 
Despite lower accuracy, DMO-based predictions broadly agree with hydro-based trends, suggesting that disk–halo relations derived from DMO runs still possess decent predictive power.
The SR relations presented here outperform previous analytic prescriptions \citep{Kravtsov13, Mo98, Jiang19}.

We try to understand the origins of these correlations.
\begin{itemize}[leftmargin=*]
\item  We find evidence of disks altering halo structure, in the sense that disk compactness influences the Einasto slope index $\alpha$.
\item Disk compactness evolves with redshift in a strongly mass-dependent way: low-mass halos become substantially more compact relative to their virial radii toward high redshift because the star-forming radius stays roughly constant while the virial radius shrinks rapidly; whereas in high-mass halos this trend reverses.
\item Disk thickness is primarily driven by halo mass and recent halo growth. Disks are thinnest in Milky Way–mass halos, thicken toward both lower and higher masses, and become systematically thicker at higher redshift. Recent and rapid halo assembly produce strong dynamical heating and thicker disks. Disk height scales tightly with disk size, so the correlations of disk height with halo spin and concentration are largely indirect through their influence on disk size.
\item Inner halo properties improve disk predictions only in hydro simulations. Inner-halo spin correlates more strongly with disk size and thickness than global spin in the hydro run because baryonic disks reshape the inner dark matter. In DMO halos, this coupling does not exist, and global halo properties more robust predictors for models based solely on $N$-body simulations.
\end{itemize}

We provide simple analytic prescriptions that use halo properties to predict a wide range of galaxy and disk properties, in \Tab{relations1} and \Tab{relations1DMO}. These relations offer computationally efficient tools for semi-analytic and empirical modelers wishing to populate halos with galaxies. Although the results here are specific to TNG50 physics, our analysis pipeline is public and modular, enabling straightforward extension to other simulations and subgrid models.

\begin{acknowledgments}
We dedicate this work to the memory of our coauthor Avishai Dekel, whose insight, generosity, and vision continue to inspire our research.
FJ and JL thank the helpful discussion with Yangyao Chen.
FJ acknowledges support by the National Natural Science Foundation of China (NSFC, 12473007) and China Manned Space Program with grant no. CMS-CSST-2025-A03.
LCH was supported by the National Science Foundation of China (12233001) and the China Manned Space Program (CMS-CSST-2025-A09).

\end{acknowledgments}

\appendix

\section{Machine learning algorithms}\label{app:MLalgorithm}
In this work, we use different ML algorithms to explore relations between halo properties and disk properties. We describe their details here. 
\subsection{Random Forest}\label{app:RF}

The RF algorithm is an ensemble-based ML method that aggregates outputs of multiple decision trees, each trained individually. A decision tree is constructed by recursively partitioning the feature space. At each node, the algorithm selects features and split points that minimize regression error. This continues until leaf nodes are reached, which give the final prediction.

\restartappendixnumbering  
\renewcommand{\thetable}{A.\arabic{table}}
\setcounter{table}{0}

\begin{table*}[ht!]
\centering
\caption{Predicted empirical relations from SR for different targets using hydro simulation with $\omega=0.7$}\label{tab:relations1}
\renewcommand{\arraystretch}{1.2}
\begin{tabularx}{0.99\textwidth}{l l L} 
\hline
\hline
Target$^*$ & $R^2$ & Model  \\ 
\hline
$M_{\star}/\Mv$ (All) & 0.78 & $(0.053 c - 1.673) e^{-0.063 M_{\rm vir,10} s} + 0.066 z_{1/2} - 1.515$ \\
SFR$\times t_{\rm H}/\Mv$ (All) & 0.58 & $0.092z_{1/2}-1.217 e^{-0.142M_{\rm vir,10}s}-1.297$ \\ 
$r_{1/2}/\Rv$ (All)$^\dagger$ & 0.65 & $\log \alpha+e^{-0.274c}+\lambda_{\rm inner}^{0.597}\log M_{\rm vir,10}+4.643\lambda zM_{\rm vir,10}^{-1.294}-1.000$ \\ 
$f_{\rm Disk}$ (All) & 0.48 & $(\lambda_{\rm inner} + 0.094) s_{\rm inner} - 0.396  e^{-0.0043  M_{\rm vir,10}c}+ 0.635$ \\ 
$f_{\rm ThinDisk}$ (All) & 0.25 & $-1.292 M_{\rm vir,10}^{0.665} c/10^4 - 0.215 \log \alpha + 0.040 \dot{M}_{\rm norm}
 + \alpha \log M_{\rm vir,10} \log (cq_{\rm inner})
 - 0.428 \lambda$ \\ 
$f_{\rm ThickDisk}$ (All) & 0.30 & $0.077 s_{\rm inner}
- 0.173 e^{-0.037 M_{\rm vir,10}}
+ \lambda_{\rm inner}^{1.209}
+ 0.251
+ 0.014 (\alpha c - 3.428)\log M_{\rm vir,10}$ \\ 
$R_{\rm 1/2, Disk}/\Rv$ (All) & 0.44 & $\lambda_{\rm inner} + (\lambda+ c^{-0.125})  (\log \alpha + 3.010) - 3.210$ \\ 
$R_{\rm 1/2, Disk}/\Rv$ (Low mass) & 0.46 &$\lambda_{\rm inner}+e^{-0.324c} + \log \alpha + \lambda^{0.203} - 1.215$\\ 
$R_{\rm 1/2, Disk}/\Rv$ (High mass) & 0.62 &$1.858  \lambda_{\rm inner} + 0.263 \log \lambda + e^{-0.231 c - 0.291} + 1.253 \log \alpha-0.626 \log (z + 1.407)$ \\ 
$R_{\rm 1/2, ThinDisk}/\Rv$ (All) & 0.40 & $2.731 \lambda+(c + 0.036 M_{\rm vir,10})^{-0.730} + \log (\alpha + \lambda_{\rm inner}) - 0.943$\\ 
$R_{\rm 1/2, ThinDisk}/\Rv$ (Low mass) & 0.37 & $(c + 0.408)^{-0.643} + \log (\lambda + 0.099) + \log (\lambda_{\rm inner} + \alpha)$ \\ 
$R_{\rm 1/2, ThinDisk}/\Rv$ (High mass) & 0.56 & $\lambda^{0.119} + \log \alpha + e^{-0.375z}(2.618  \lambda_{\rm inner} + 0.327)  - 0.231 \log c - 1.323$\\ 
$R_{\rm 1/2, ThickDisk}/\Rv$ (All)& 0.60 & $\log (\lambda_{\rm inner} + \alpha) - 0.672 \log (0.043 M_{\rm vir,10} + c) - 0.188$ \\ 
$R_{\rm 1/2, ThickDisk}/\Rv$ (Low mass) & 0.59 & $0.722 z_{1/2} \lambda + \log (\lambda_{\rm inner} + \alpha)  - 0.512  \log c - 0.416$ \\ 
$R_{\rm 1/2, ThickDisk}/\Rv$ (High mass) & 0.68 & $\lambda^{0.421} -0.303  \log c - 0.069 z_{1/2}  \log (0.105  M_{\rm vir,10}) + \log (\alpha \lambda_{\rm inner} + 0.162)$\\ 
$Z_{\rm 1/2, Disk}/\Rv$ (All) & 0.79 & $\lambda_{\rm inner} - (1.681  \alpha + 0.378)\log c  + \log \alpha - 0.386  z  \log (0.030  M_{\rm vir,10} + 0.750) + 0.112  \log n_1$ \\ 
$Z_{\rm 1/2, Disk}/\Rv$ (Low mass) & 0.74 & $\alpha + 0.097  (\lambda_{\rm inner} - 0.123) M_{\rm vir,10} + n_1^{0.087} + z  \lambda - 2.052 c^{0.093}$\\ 
$Z_{\rm 1/2, Disk}/\Rv$ (High mass) & 0.70 & $1.967  \lambda_{\rm inner} + 0.241  (\alpha - 0.013) c - 2.403 + e^{z_{1/2} (\lambda - 0.147)} - 10.667  \alpha \log (0.186 c) + \log \alpha$ \\ 
$Z_{\rm 1/2, ThinDisk}/\Rv$ (All) & 0.72 & $(\lambda_{\rm inner} - 0.253)  \log M_{\rm vir,10} + e^{-0.191c}  - 0.082  \alpha^{-0.753} + z  \lambda- 1.912$\\ 
$Z_{\rm 1/2, ThinDisk}/\Rv$ (Low mass) & 0.70 & $\alpha + (z \lambda)^{0.694} - 1.939  c^{0.087} + 0.109  (\lambda_{\rm inner} - 0.128)  M_{\rm vir,10}$\\ 
$Z_{\rm 1/2, ThinDisk}/\Rv$ (High mass) & 0.58 & $\log \alpha -0.416  \log (c + 1.177) + z  (M_{\rm vir,10}^{-0.447} - 0.227) + 1.965  (\lambda_{\rm inner} + \lambda) + 0.693  - 1.582$\\ 
$Z_{\rm 1/2, ThickDisk}/\Rv$ (All) & 0.78 & $0.593  \log \alpha +\lambda_{\rm inner}+ n_1^{0.122}  e^{p_{\rm inner}} - 0.532  \log c  - 0.220 z_{1/2}  \log (0.024  M_{\rm vir,10} + 0.778) - 1.463
$ \\ 
$Z_{\rm 1/2, ThickDisk}/\Rv$ (Low mass) & 0.71 & $0.075 M_{\rm vir,10}  (\alpha + \lambda_{\rm inner} - 0.259) - 0.508  \log c + 5.870  (n_1^{0.191} - 0.311)
$ \\ 
$Z_{\rm 1/2, ThickDisk}/\Rv$ (High mass) & 0.69 & $ 2.799 \lambda_{\rm inner} - \log c  (0.617  \log \alpha + 1.194) + 1.324  (\log \alpha - 0.399)  + z_{1/2}  (M_{\rm vir,10}^{-0.650} - 0.145)
$\\ 
\hline
\end{tabularx}
\begin{flushleft}
\ \ $^*$ All targets are in base-10 logarithm except disk mass fractions.\\
\ \ $^\dagger$ $\omega=0.9$ is used for this equation.
\end{flushleft}
\end{table*}

\begin{table*}[ht!]
\centering
\caption{Predicted empirical relations from SR for different targets using DMO simulation with $\omega=0.7$}\label{tab:relations1DMO}
\renewcommand{\arraystretch}{1.2}
\begin{tabularx}{0.99\textwidth}{l l L} 
\hline
\hline
Target$^*$ & $R^2$ & Model  \\ 
\hline
$M_{\star}/\Mv$ (All) & 0.73 & $0.081 z_{1/2} - 1.375 e^{-0.089 M_{\rm vir,10} s}  +0.042 c - 1.900$ \\
SFR$\times t_{\rm H}/\Mv$ (All) & 0.54 & $0.101z_{1/2}+e^{-0.146M_{\rm vir,10}s}(0.014M_{\rm vir,10}^{-2.337}-1.334)-1.296$ \\ 
$r_{1/2}/\Rv$ (All) & 0.54 & $\log (\lambda + 0.162) \log (M_{\rm vir,10} + c) + \log (z + s) - 1.474 + e^{-1.272M_{\rm vir,10} + 0.642} + \lambda_{\rm inner} + (0.643-0.201  z^{0.463})(\log M_{\rm vir,10} + 0.374)$ \\ 
$f_{\rm Disk}$ (All) & 0.43 & $0.160  q_{\rm inner}+ c^{0.476} M_{\rm vir,10}(M_{\rm vir,10} + 35.67)^{-1.31} + 0.172$ \\ 
$f_{\rm ThinDisk}$ (All) & 0.28 & $0.077(q_{\rm inner}+\log z_{1/2})+-0.284e^{-0.010cM_{\rm vir,10}}+0.310$ \\ 
$f_{\rm ThickDisk}$ (All) & 0.20 & $0.477\lambda_{\rm inner}+(0.052q_{\rm inner}-0.092)\log_{\rm 10}M_{\rm vir, 10}-0.243e^{-0.063sM_{\rm vir,10}}+0.382$ \\ 
$R_{\rm 1/2, Disk}/\Rv$ (All) & 0.41 & $\log M_{\rm vir,10} (\lambda^{0.251} - 0.339) + z  (\lambda + 0.088) + e^{-0.021  c -0.007 z M_{\rm vir,10}+0.045} + 0.343  q - 2.696$ \\ 
$R_{\rm 1/2, Disk}/\Rv$ (Low mass) & 0.29 & $\lambda^{-0.084}  (\lambda_{\rm inner} - 1.122) + e^{M_{\rm vir,10}\lambda - 0.090}  \log (0.587 z + q)$\\ 
$R_{\rm 1/2, Disk}/\Rv$ (High mass) & 0.27 &$-0.387 \log M_{\rm vir,10}  z  e^{\dot{M}_{\rm norm}} + \log \lambda + 0.328 (z^{1.101} + q) - 3.663 \lambda  \log M_{\rm vir,10}$ \\ 
$R_{\rm 1/2, ThinDisk}/\Rv$ (All) & 0.37 & $n_1^{0.087} + (\lambda^{0.344} + \lambda_{\rm inner})  \log M_{\rm vir,10} + (\lambda + 0.578)  (\log q + z) - 0.359  (z + 1.284)  (M_{\rm vir,10} + c)^{0.177} - 1.106
$\\ 
$R_{\rm 1/2, ThinDisk}/\Rv$ (Low mass) & 0.33 & $(\log z_{1/2} + 0.098  M_{\rm vir,10})  (-0.063 z + \lambda_{\rm inner})  + 0.335  \log (q  \lambda) + z  (\lambda + 0.146)- 0.906$ \\ 
$R_{\rm 1/2, ThinDisk}/\Rv$ (High mass) & 0.29 & $0.486  \log (\lambda  q) + (M_{\rm vir,10}^{-0.221} - 0.969)  (0.992 + z)^{0.332}
$\\ 
$R_{\rm 1/2, ThickDisk}/\Rv$ (All) & 0.52 & $\lambda^{0.193}  (z + 0.519)^{0.396} - 1.344 z  (M_{\rm vir,10}^{-0.971} + 0.067)^{-3.987}/10^{-5}  + \lambda_{\rm inner} - 2.141
$ \\ 
$R_{\rm 1/2, ThickDisk}/\Rv$ (Low mass) & 0.48 & $\lambda_{\rm inner} + \lambda^{0.268} - 0.005  M_{\rm vir,10}  z - 0.020  c + \log (z + 2.054 q) - 2.153$ \\ 
$R_{\rm 1/2, ThickDisk}/\Rv$ (High mass) & 0.38 & $0.481  (z + \log \lambda)-(0.375 z + 0.101)  (\log M_{\rm vir,10} + \dot{M}_{\rm norm}) -0.940
$\\ 
$Z_{\rm 1/2, Disk}/\Rv$ (All) & 0.68 & $\lambda^{0.435} -0.446  \log c + (p_{\rm inner} + z^{0.596})  (-0.237 M_{\rm vir,10}^{0.224} + 0.534) - 2.219 
$ \\ 
$Z_{\rm 1/2, Disk}/\Rv$ (Low mass) & 0.66 & $\lambda^{0.302} - 0.461  \log c + (z + 0.208)^{0.515}  (-0.012  M_{\rm vir,10} + 0.350) - 2.318$\\ 
$Z_{\rm 1/2, Disk}/\Rv$ (High mass) & 0.23 & $1.931 \lambda_{\rm inner} + (z_{1/2} -0.082 c )  (0.105 M_{\rm vir,10})^{-0.938} + 0.158 z_{1/2}  \log \lambda  + 0.389 s- 2.600$ \\ 
$Z_{\rm 1/2, ThinDisk}/\Rv$ (All) & 0.65 & $(\lambda + 0.714)  (z + 1.932) - (0.424 z + 0.184)  \log (M_{\rm vir,10} + 3.137  z_{1/2})  - 3.630
$\\ 
$Z_{\rm 1/2, ThinDisk}/\Rv$ (Low mass) & 0.65 & $-0.813  e^{-0.326 z} + 0.095 z_{1/2}  M_{\rm vir,10}  (\lambda - 0.086) - 0.282  \log c - 1.214$\\ 
$Z_{\rm 1/2, ThinDisk}/\Rv$ (High mass) & 0.26 & $(M_{\rm vir,10}^{-0.4900863} - 0.174) 3.048 z  - 0.088 z^{1.898} [\log (\alpha + 0.209) + \text{PC2}] + 0.331 (q_{\rm inner} + \log \lambda)- 2.281$\\ 
$Z_{\rm 1/2, ThickDisk}/\Rv$ (All) & 0.70 & $(z^{0.736} + s)  (\lambda + 0.291) -0.379 z_{1/2} \log (0.042 M_{\rm vir,10} + 1.064)  - 2.412
$ \\ 
$Z_{\rm 1/2, ThickDisk}/\Rv$ (Low mass) & 0.66 & $0.206  (s + \log \lambda) + (0.315-0.011  M_{\rm vir,10})  z^{0.562} - 0.426  \log c - 1.669$ \\ 
$Z_{\rm 1/2, ThickDisk}/\Rv$ (High mass) & 0.38 & $\lambda_{\rm inner}+0.155  z_{1/2}  \log \lambda + 9.116z_{1/2}  c^{-0.272}  M_{\rm vir,10}^{-0.880603} - 2.322
$\\ 
\hline
\end{tabularx}
\begin{flushleft}
\ \ $^*$ All targets are in base-10 logarithm except disk mass fractions.
\end{flushleft}
\end{table*}


Although decision trees are interpretable, they often overfit. RF mitigates this by adding controlled randomness where each tree is trained on a bootstrap sample, and at each split only a random subset of features is considered. After all trees are built, RF averages their outputs, producing more robust predictions and reducing overfitting.

We use the RF regressor from \texttt{scikit-learn} \citep{Pedregosa11}. The hyperparameters tuned in the module are listed below while other hyperparameters remain default settings:
\begin{itemize}
\item \texttt{n\_estimators}: the number of decision trees in the RF, taken from [200, 300, 400, 500, 600, 800, 1000, 1200].
\item \texttt{max\_depth}: the maximum depth of each tree, taken from [7, 10, 13, 16, 19, 22, 25].
\item \texttt{min\_samples\_leaf}: the minimum number of samples required at a leaf node, taken over [10, 20, 50, 70, 95, 110].
\item \texttt{max\_features}: the maximum number of features to consider when looking for the best split, with options from [\texttt{sqrt}, \texttt{log2}, \texttt{None}]. Specifically, \texttt{sqrt} uses the square root of the total number of features, \texttt{log2} uses the base-2 logarithm, and \texttt{None} uses all available features.
\end{itemize}

We use the \texttt{GridSearchCV} module from \texttt{scikit-learn} to tune the hyperparameters defined above. This module applies K-fold cross-validation, where the training set is split into K folds: one fold is used for validation while the remaining K-1 folds are used for training, repeated K times. For each hyperparameter combination, performance is evaluated using the chosen metric, and the best set of parameters is selected. In our experiments, we use 5-fold cross-validation with negative mean squared error (\texttt{neg\_mean\_squared\_error}) as the scoring metric.

\subsection{Shapley Additive Explanations Values}\label{app:SHAP}



To interpret regression results from the complex tree structures in RF, we use SHAP \citep{Lundberg17}, a game-theoretic method. SHAP evaluates how much each feature contributes to a prediction by comparing the model’s output with and without that feature across many possible feature combinations. A SHAP value is the weighted average of these marginal contributions, capturing both the direction and strength of a feature’s effect.

\begin{figure*}	
\centering
\includegraphics[width=0.77\textwidth]{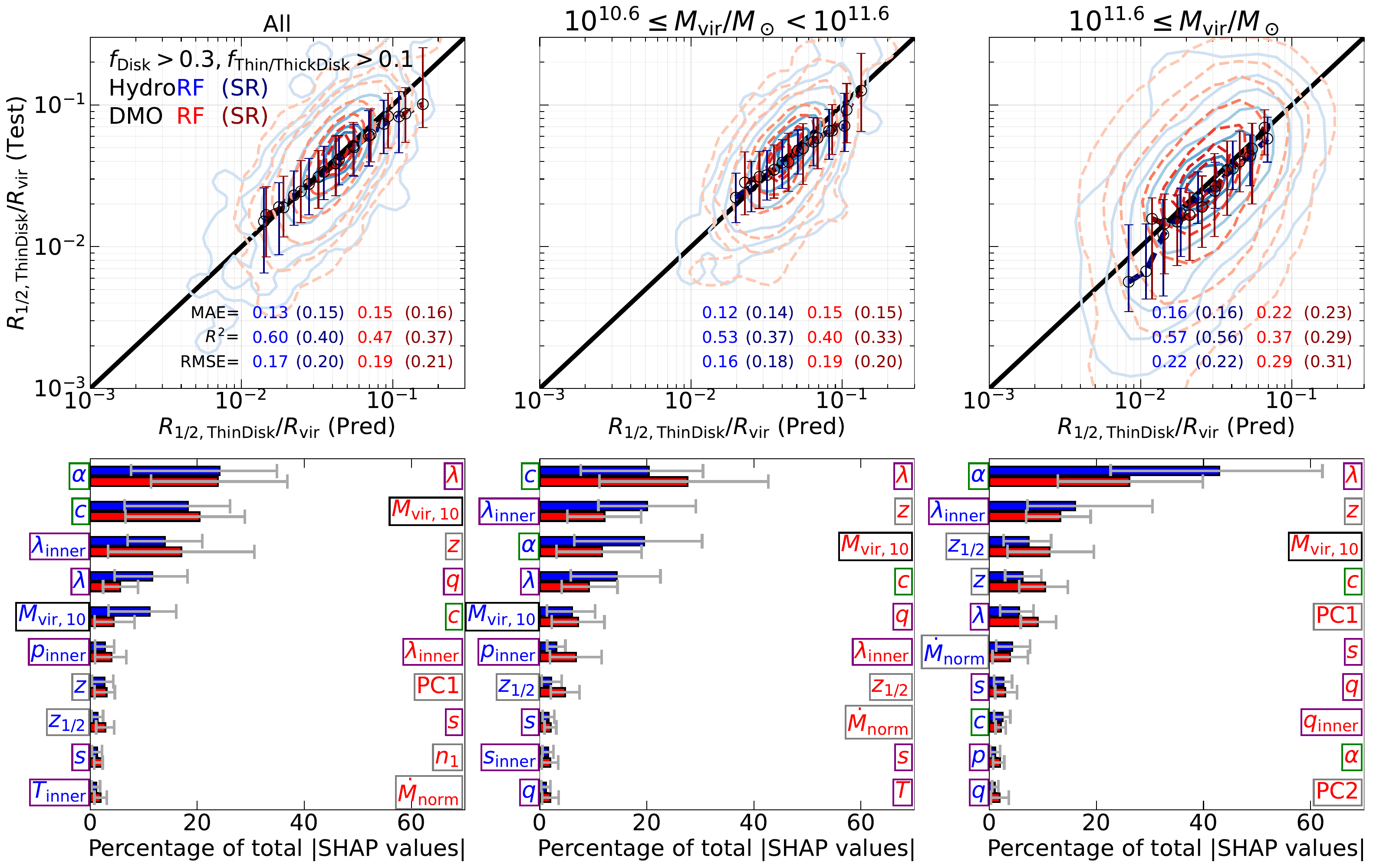}
    \caption{\textbf{Predictions for thin disk compactness $R_{\rm 1/2,ThinDisk}/R_{\rm vir}$ from ML algorithms.} The colors and styles for lines and symbols are similar to \Fig{DiskSize}, except the samples are further limited by $f_{\rm ThinDisk}>0.1$ and $f_{\rm ThickDisk}>0.1$.}
    \label{fig:ThinDiskSize}
\end{figure*}

A positive SHAP value indicates that a feature increases the prediction relative to a baseline, while a negative value indicates a decrease. For each data point, the sum of all SHAP values equals the difference between the model’s prediction and the baseline. The absolute values show how strongly each feature influences the prediction. By aggregating SHAP values across data points, we can identify and rank the most influential features. However, SHAP does not measure model accuracy or reveal direct correlations among features.

\subsection{Symbolic Regression}\label{app:SR}
For finding empirical and interpretable equations from the most important features identified by SHAP, we apply the SR \citep[e.g.][]{Koza92} equation
search algorithm \texttt{PySR} \citep{Cranmer23}. 

\texttt{PySR} is based on genetic programming, which searches for multiple candidate equations (populations) in parallel rather than just a single solution. Each candidate equation is represented as a tree structure, where the nodes are mathematical operators and the leaves are variables or constants.

The algorithm begins by randomly initializing populations of simple equations using the given input features and operators. It then evaluates the loss of each equation on the data and iteratively improves them through evolutionary steps including mutation (changing an operator or replacing a subtree), crossover (swapping branches between equations), and selection (keeping the equations that minimize the loss). Afterward, \texttt{PySR} applies simplification rules and optimizes constants within the equations.

Once all populations complete one evolutionary cycle (a generation), migration occurs which exchange a fraction of equations between populations or replace them with top-performing candidates from the overall pool. This promotes diversity and helps avoid getting stuck in local optima.

The \texttt{PySR} configuration in this work is implemented on a large-scale setup with 150 populations, each containing 27 individuals, allowing the algorithm to run for up to 300 iterations with 500 cycles per iteration, thereby ensuring thorough exploration of the expression space. The model is restricted to the binary operators multiplication ($\times$), plus ($+$), and power (\ $ \widehat{} $\ ), and the unary operators exponential ($\exp$) and base-10 logarithm ($\log $). Variable complexity is set to 3, while constant and operator complexities are set to 1. Strict complexity constraints are enforced, including a maximum expression size of 100 and a depth limit of 8. Furthermore, nested operations of exponential, base-10 logarithm, and power are prohibited to avoid overly complex expressions. Weight optimization is set to 0.001, with higher values resulting in more frequent optimization steps. All other hyperparameters are left at their default settings.

\begin{figure*}	
\centering
\includegraphics[width=0.77\textwidth]{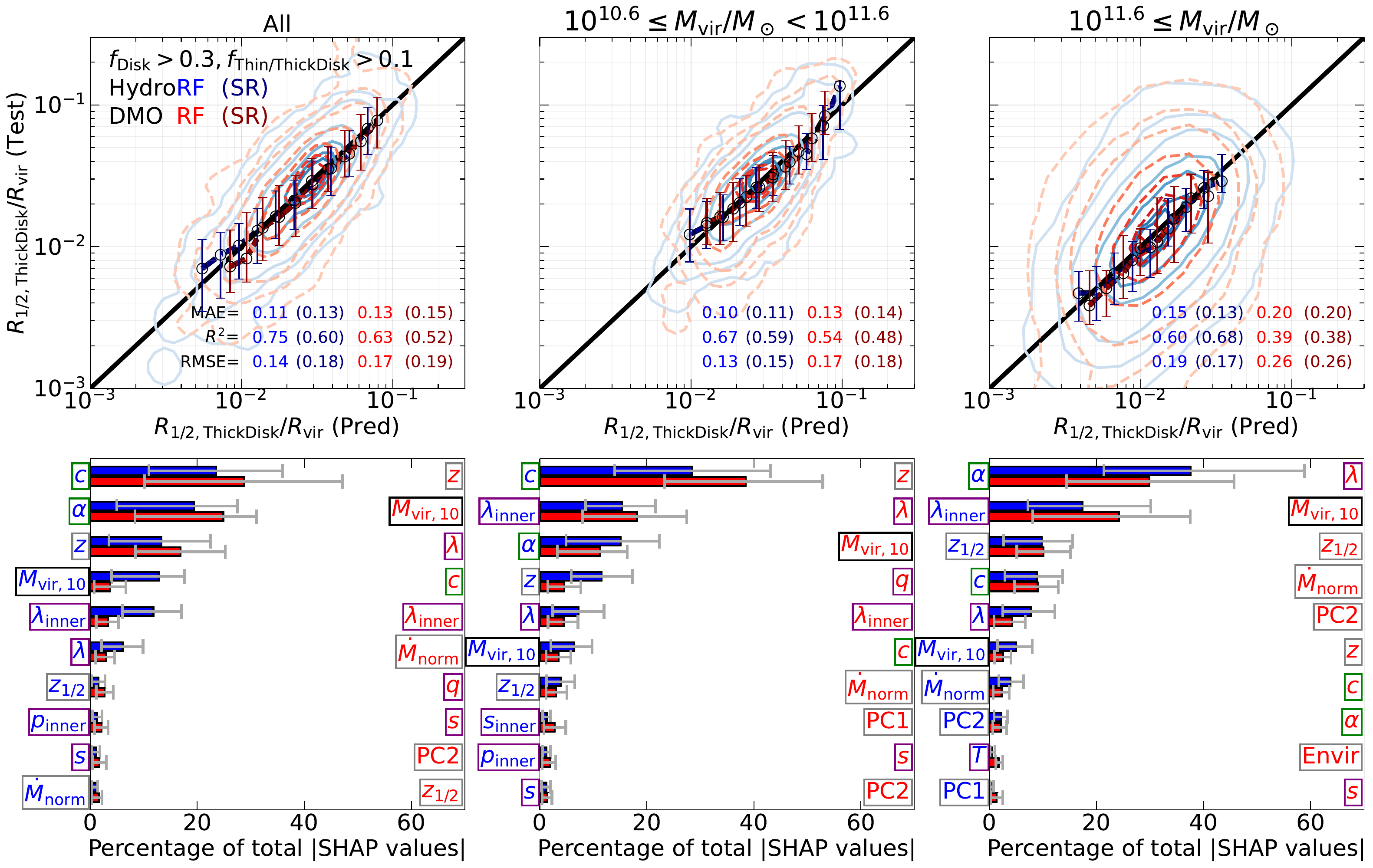}
    \caption{\textbf{Performance of regression models for thick disk compactness $R_{\rm 1/2,ThickDisk}/R_{\rm vir}$ from ML algorithms.} The colors and styles for lines and symbols are similar to \Fig{DiskSize}, except the samples are further limited by $f_{\rm ThinDisk}>0.1$ and $f_{\rm ThickDisk}>0.1$.}
    \label{fig:ThickDiskSize}
\end{figure*}


\section{Empirical relations from symbolic regression}\label{app:SRrelations}
In this section, we provide our empirical relations from SR using $w=0.7$ in \Tab{relations1} and \Tab{relations1DMO} for hydro and DMO simualtion, respectively. Other equations returned by symbolic regression are provided in \href{https://github.com/JinningLianggithub/Gal-halo_Symbolic-regression}{https://github.com/JinningLianggithub/Gal-halo\_Symbolic-regression}.

\section{Thin and thick disks}
In this section, we show the RF and SR results for both thin and thick disks, which are separated using the morphological decomposition method (see Paper I for details)
In addition to applying the total disk mass fraction cut of $f_{\rm Disk}>0.3$, we further require a thin disk mass fraction of $f_{\rm ThinDisk}>0.1$ and a thick disk mass fraction of $f_{\rm ThickDisk}>0.1$.

\begin{figure*}	
\centering
\includegraphics[width=0.77\textwidth]{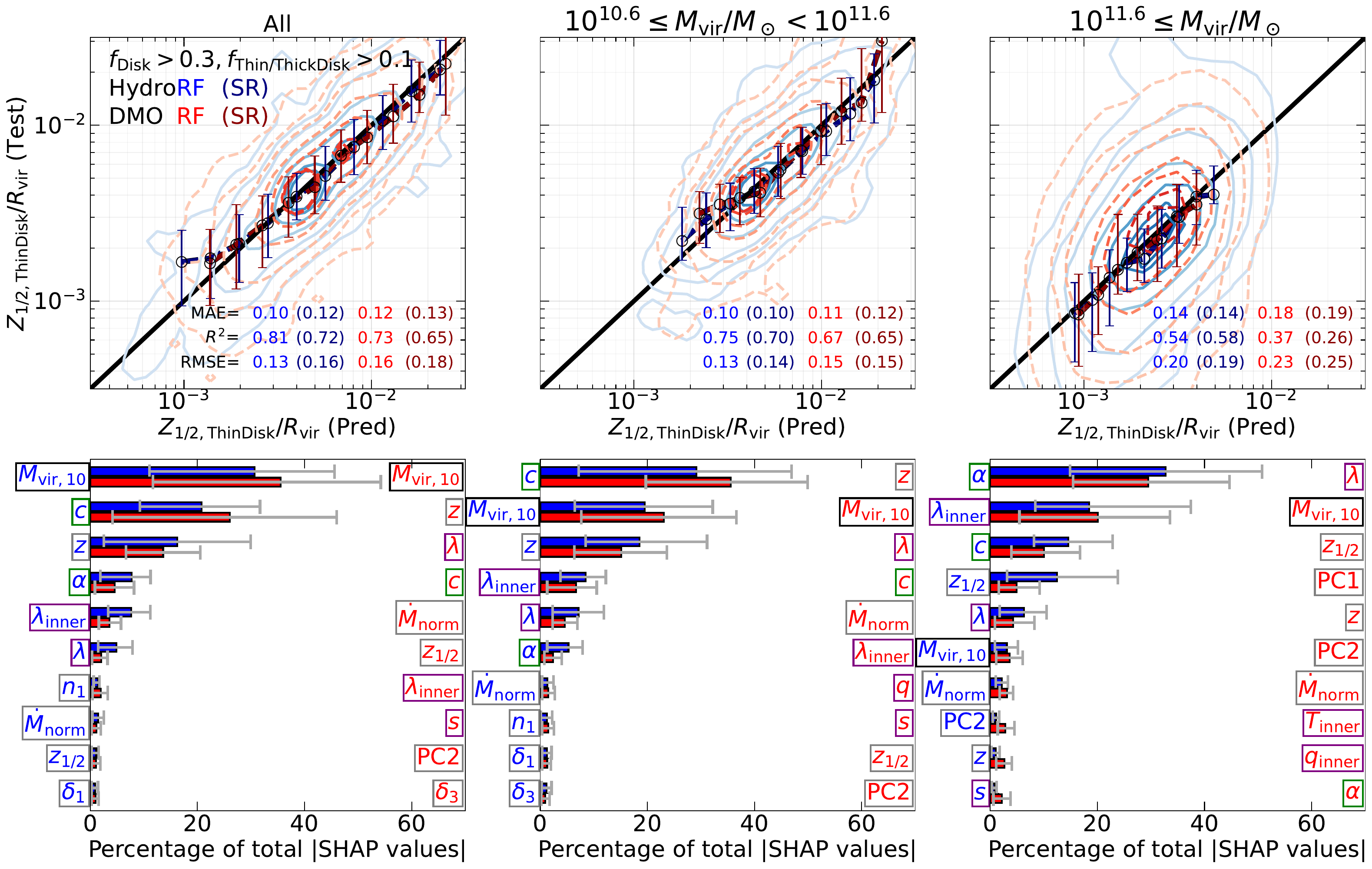}
    \caption{\textbf{Performance of regression models for thin disk thickness $Z_{\rm 1/2,ThinDisk}/R_{\rm vir}$.} The samples are limited to $f_{\rm ThinDisk}>0.1$ and $f_{\rm ThickDisk}>0.1$. Everything else follows \Fig{DiskSize}.}
    \label{fig:ThinDiskHeight}
\end{figure*}
\subsection{Sizes}
\Fig{ThinDiskSize} and \Fig{ThickDiskSize} present predictions for the compactness of thin and thick disks, respectively. 

{\bf Model accuracy:}
The thin disk compactness is predicted with accuracy comparable to that of the total disk, yielding $R^2=0.60$ (0.47) from RF and $R^2=0.40$ (0.37) from SR, using hydro (DMO) dataset. 
The thick disk is even more predictable, with $R^2=0.75$ (0.63) from RF, $R^2=0.60$ (0.52) from SR, using hydro (DMO) dataset.

{\bf Feature importance:}
For hydro dataset, the most important halo properties for both thin and thick disk compactness are $\alpha$, $c$, $M_{\rm vir,10}$, $\lambda$, $\lambda_{\rm inner}$, and $z$. These are nearly the same as those for the total disk. 
Similarly, for DMO halos, the shared important properties are $z$, $M_{\rm vir}$, $\lambda$, $c$.

{\bf Symbolic regression:}
While the SR models for thin disk compactness reach accuracy levels similar to those of the total disk, the thick disk prediction are noticeably better. The full analytic forms of the SR relations are listed in \Tab{relations1} and \Tab{relations1DMO}.

{\bf Mass dependence:}
For both thin and thick disks, the $R^2$ values are lower in each mass bin than for the full sample, for both RF and SR. The set of important features remains broadly similar. They are $\alpha$, $\lambda_{\rm inner}$, $c$ and $\lambda$ for the thin and thick disks using hydro dataset. In the DMO dataset, common important features include $\lambda$, $M_{\rm vir,10}$, $z$, $c$ for the thin disk. For the thick disk, low-mass halos also show $\lambda$, $M_{\rm vir,10}$, $z$ are important. Differently, high-mass halos show more formation and environmental parameters are important, which include $z_{1/2}$, $\dot{M}_{\rm norm}$ and PC2,

\subsection{Scale heights}
\begin{figure*}	
\centering
\includegraphics[width=0.77\textwidth]{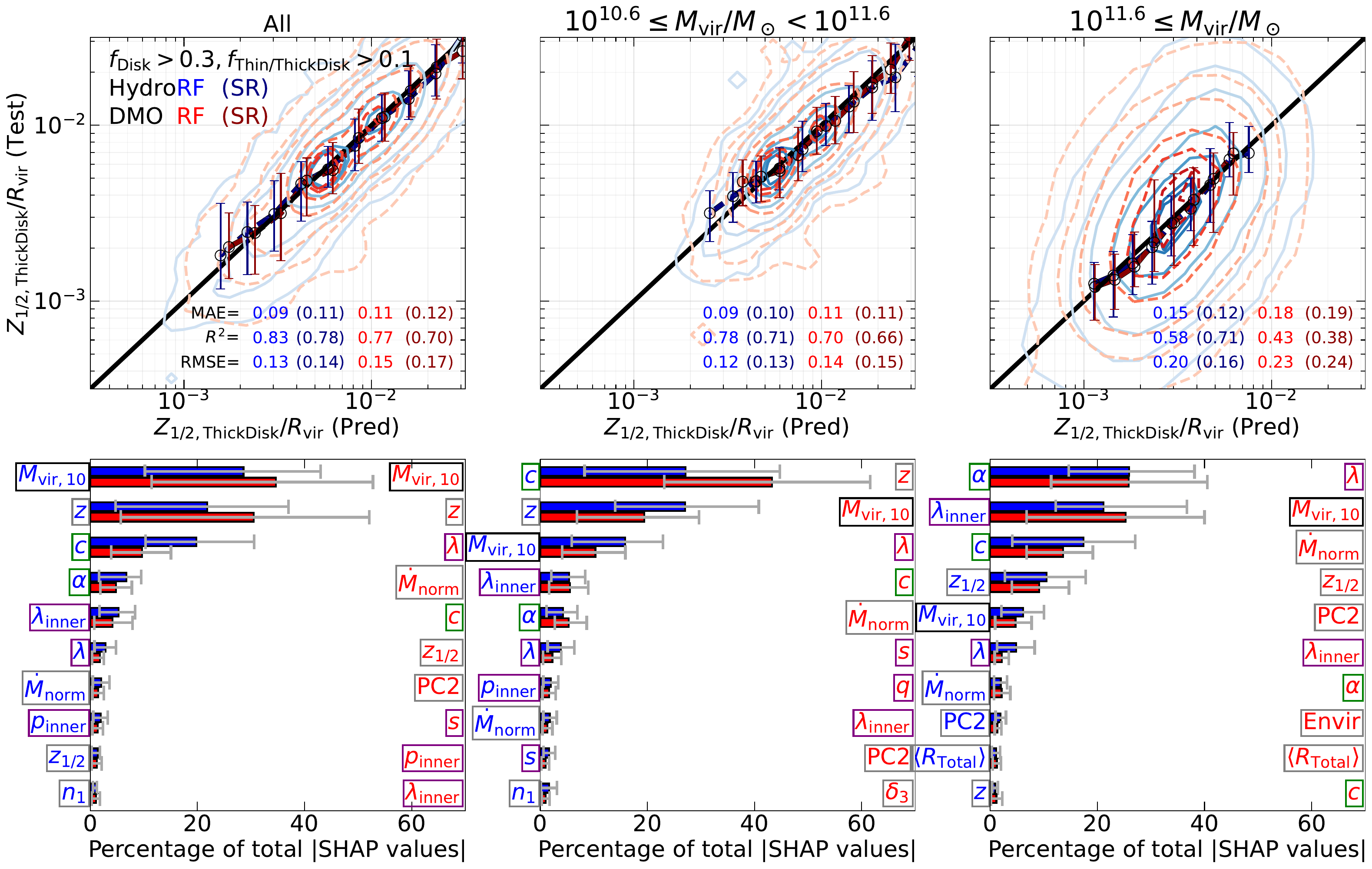}
    \caption{\textbf{Performance of regression models for thick disk thickness $Z_{\rm 1/2,ThickDisk}/R_{\rm vir}$.} The samples are limited to $f_{\rm ThinDisk}>0.1$ and $f_{\rm ThickDisk}>0.1$. Everything else follows \Fig{DiskSize}.}
    \label{fig:ThickDiskHeight}
\end{figure*}

The prediction for thickness of thin and thick disks are presented in \Fig{ThinDiskHeight} and \Fig{ThickDiskHeight}, respectively. 

{\bf Model accuracy:}
Similar to the total disk, the prediction accuracies for thin and thick disk thicknesses are noticeably higher than those for their compactness, for both RF and SR. The predictions for thin disk thickness achieve a $R^2=0.81$ (0.73) from RF and $R^2=0.72$ (0.65) from SR, using hydro (DMO) dataset. Similarly, the thick disk achieve $R^2=0.83$ (0.77) from RF, $R^2=0.78$ (0.70) from SR, using hydro (DMO) dataset.

{\bf Feature importance:}
Using the hydro dataset, $M_{\rm vir,10}$, $\alpha$, $c$, $\lambda$, $\lambda_{\rm inner}$, and $z$ remain the most important halo properties for the disk thickness of thin and thick disks, the same as their compactness. 
With the DMO dataset,  $z$, $M_{\rm vir,10}$, $\lambda$, $c$, and $\dot{M}_{\rm norm}$ are the dominant halo parameters for the predictions of thin and thick disk thickness. The appear of $\dot{M}_{\rm norm}$ is slightly different from their disk compactness, which potentially highlights the disk heating by accretion.

{\bf Symbolic regression:}
The SR prediction accuracy for thin and thick disk thicknesses is similar to that for the total disk. The full analytic SR expressions are provided in \Tab{relations1} and \Tab{relations1DMO}. Notably, the SR models capture additional formation and environmental properties including $z_{1/2}$ and $n_1$ in the relations.

{\bf Mass dependence:}

Using the hydro dataset, the key halo properties within individual mass bins remain broadly consistent with those of the full sample for both disk types. In both low  and high mass halos, halo mass ranks lower in importance. In the high-mass halos, $z_{1/2}$ emerges as a more important parameter, replacing $z$. With the DMO dataset, low-mass halos show the same dominant parameters as in the full sample for both disks. In high-mass halos, both disk types show increased importance in formation and environmental properties.

\subsection{Mass fraction}
\begin{figure*}	
\centering
    \includegraphics[width=0.5\textwidth]{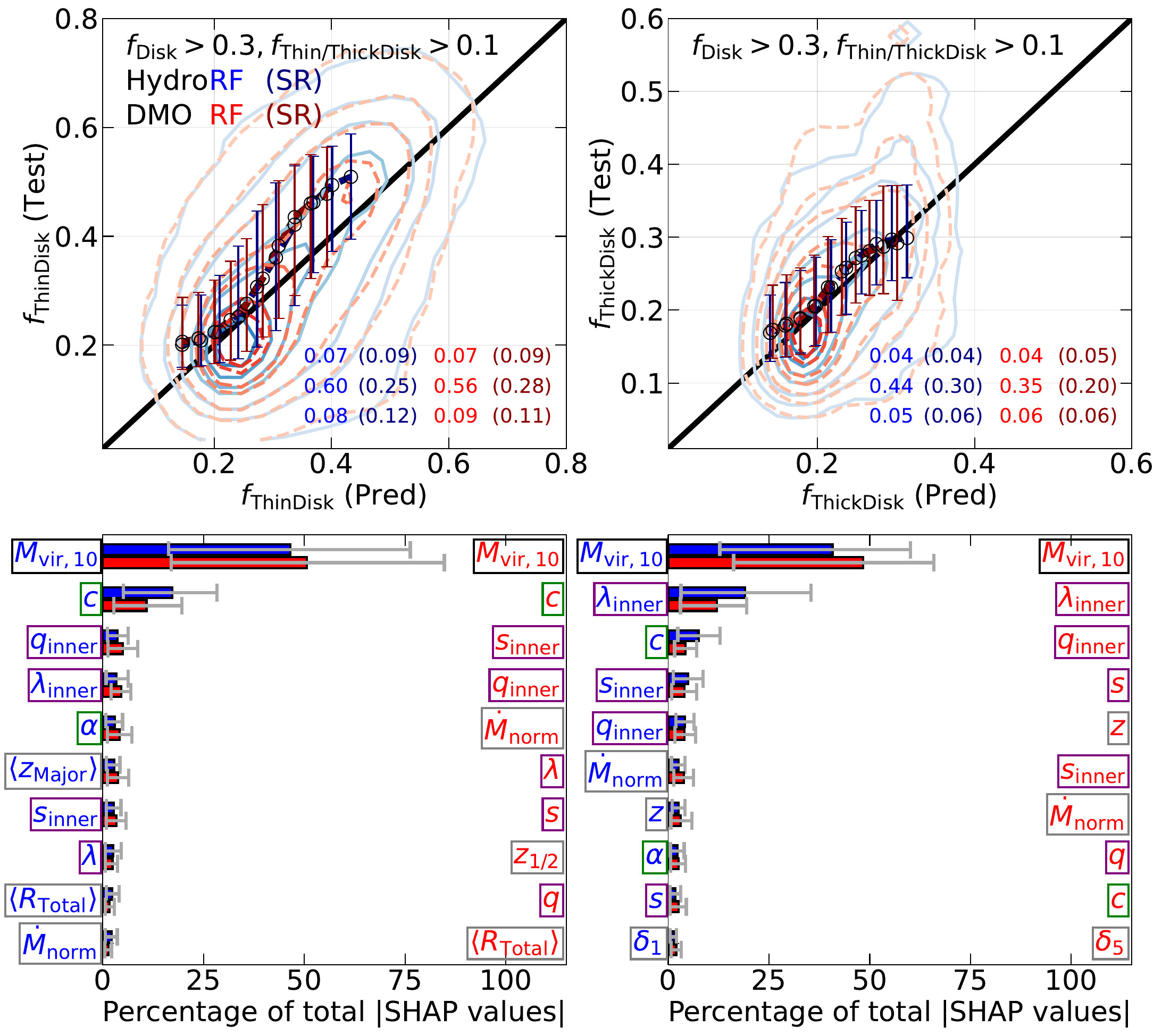}
    \caption{\textbf{Performance of regression models for thin and thick disk mass fractions.} The left and right columns present the results for the thin ($f_{\rm ThinDisk}$), and thick disk ($f_{\rm ThickDisk}$) mass fractions, respectively. The sample is limited to $f_{\rm Disk} > 0.3$, $f_{\rm ThinDisk} > 0.1$, and $f_{\rm ThickDisk} > 0.1$.  Everything else follows \Fig{DiskSize}.}
    \label{fig:f_ThinThick}
\end{figure*}

The predictions for the mass fractions of thin and thick disks are shown in \Fig{f_ThinThick}.

{\bf Model accuracy:}
As with the total disk, the sample is not divided into different halo mass bins. The prediction accuracy for thin disk mass fraction is higher than that for the total disk, reaching $R^2=0.60$ (0.56) for the hydro (DMO) dataset. For both datasets, the accuracy for thin disks is about 0.2 higher than for thick disks.
The SR models exhibit lower accuracy than the RF models, with deficits of roughly 0.15–0.3 for both thin and thick disk mass fractions.

{\bf Feature importance:}
Except for halo mass and one additional key property ($c$ for the thin disk and $\lambda_{\rm inner}$ for the thick disk), the remaining halo properties all have rather weak importance.
{\bf Symbolic regression:}
In the SR relation, the halo mass and second property above are captured. Inner shape parameters (e.g. $q_{\rm inner}$ and $s_{\rm inner}$) and formation parameters (e.g. $\dot{M}_{\rm norm}$ and $z_{1/2}$) are also incorporated. The explicit forms of the SR relations are provided in \Tab{relations1} and \Tab{relations1DMO}.

\bibliography{sample701}{}
\bibliographystyle{aasjournalv7}



\end{document}